\documentclass[twocolumn,showpacs,preprintnumbers,nofootinbib,aps,prd,superscriptaddress,notitlepage,10pt]{revtex4-1}

\usepackage{amsfonts,amsmath,amssymb,mathrsfs}
\usepackage{graphicx,epsf, epsfig, amssymb}
\usepackage[linktocpage]{hyperref}
\usepackage{cleveref}
\usepackage{color}
\usepackage[usenames,dvipsnames]{xcolor}
\usepackage{booktabs,array}
\usepackage[british]{babel}
\usepackage[utf8]{inputenc}

\definecolor{coolblack}{rgb}{0.0, 0.18, 0.39}
\definecolor{darkred}{rgb}{0.5,0,0}
\definecolor{darkgreen}{rgb}{0,0.5,0}
\definecolor{darkblue}{rgb}{0,0,0.5}
\hypersetup{colorlinks=true, citecolor=darkblue, linkcolor=darkblue,
urlcolor = darkblue}

\newcommand{\tn}{\textnormal}

\begin{document}

\title{Neutron stars in Horndeski gravity}

\author{Andrea Maselli}
\email{andrea.maselli@uni-tuebingen.de}
\affiliation{Theoretical Astrophysics, Eberhard Karls University of Tuebingen,
Tuebingen 72076, Germany}

\author{Hector O. Silva}
\email{hosilva@phy.olemiss.edu}
\affiliation{Department of Physics and Astronomy, The University of Mississippi, University, Mississippi 38677, USA}

\author{Masato Minamitsuji}
\email{masato.minamitsuji@ist.utl.pt}
\affiliation{CENTRA, Departamento de F\'isica, Instituto Superior
T\'ecnico, Universidade de Lisboa, Avenida Rovisco Pais 1,
1049 Lisboa, Portugal}

\author{Emanuele Berti}
\email{eberti@olemiss.edu}
\affiliation{Department of Physics and Astronomy, The University of
  Mississippi, University, Mississippi 38677, USA}
\affiliation{CENTRA, Departamento de F\'isica, Instituto Superior
T\'ecnico, Universidade de Lisboa, Avenida Rovisco Pais 1,
1049 Lisboa, Portugal}

\pacs{04.50.Kd, 97.60.Jd}

\date{\today}

\begin{abstract}
  Horndeski's theory of gravity is the most general scalar-tensor
  theory with a single scalar whose equations of motion contain at
  most second-order derivatives. A subsector of Horndeski's theory
  known as ``Fab Four'' gravity allows for dynamical self-tuning of
  the quantum vacuum energy, and therefore it has received particular
  attention in cosmology as a possible alternative to the $\Lambda$CDM
  model. Here we study compact stars in Fab Four gravity, which
  includes as special cases general relativity (``George''),
  Einstein-dilaton-Gauss-Bonnet gravity (``Ringo''), theories with a
  nonminimal coupling with the Einstein tensor (``John''), and theories
  involving the double-dual of the Riemann tensor (``Paul''). We
  generalize and extend previous results in theories of the John class
  and were not able to find realistic compact stars in theories
  involving the Paul class.
\end{abstract}

\maketitle

\section{Introduction}


The most recent cosmological observations are consistent with standard
cosmological models built on general relativity (GR), but they imply
the presence of a mysterious late-time acceleration phase.  The
late-time acceleration can be interpreted as due to the existence of
new particle sectors beyond the Standard Model, or explained by
assuming that GR itself is modified on cosmological scales.  Modified
gravity models differ widely in their physical motivations, but many
of them can be reformulated in terms of scalar-tensor theories of
gravitation; i.e., they are mathematically equivalent to a
gravitational theory whose degrees of freedom are the metric
$g_{\mu\nu}$ and one or more scalar fields $\phi$.  Many of the
simplest dark energy or modified gravity models -- including the
standard $\Lambda$CDM model -- are plagued by the cosmological
constant problem (i.e., the problem of fine-tuning the potentially
huge quantum vacuum energy against the small value of the observed
cosmological constant).  However some scalar-tensor theories allow for
a ``dynamical self-tuning mechanism'' in which the effects of the
cosmological constant may be compensated within the scalar field
sector, so that they do not appear in the metric, by relaxing the
assumptions of Weinberg's no-go theorem \cite{Weinberg:1988cp}.  Here
we will focus on one such model, called ``Fab Four'' gravity in the
literature, which is a special case of Horndeski's theory.

\subsection{Horndeski's theory}
\label{sec:Horndeski}

Realistic models of dark energy or modified gravity must at the very
least pass the stringent experimental constraints on deviations from
GR~\cite{Will:2005va,Berti:2015itd} and be theoretically viable. In
particular, they must be free of the so-called ``Ostrogradski
ghost''~\cite{Woodard:2015zca}.  Several studies led to the conclusion
that the most general models with a single additional scalar degree of
freedom compatible with these requirements correspond to the
scalar-tensor theory formulated by Horndeski about 40 years ago,
whose equations of motion contain at most second-order
derivatives~\cite{Horndeski:1974wa}.  It was
shown~\cite{Kobayashi:2011nu} that Horndeski's theory is equivalent to
the generalization of a scalar field theory with Galilean shift
symmetry in flat spacetime to curved spacetime~\cite{Deffayet:2009mn},
whose action reads
\begin{equation}
S=\sum_{i=2}^{5}\int d^{4}x\sqrt{-g}{\cal L}_i\,,
\label{eq:action}
\end{equation}
where
\begin{subequations}
\label{eq:lagrangean}
\begin{align}
{\cal L}_2&=G_2\,,\label{g2}\\
{\cal L}_3&=-G_{3}\square\phi\,,\\
{\cal L}_4&=G_{4}R+G_{4\tn{X}}\left[(\square\phi)^2-\phi_{\mu\nu}^2\right]\,,\\
{\cal L}_5&=G_{5}G_{\mu\nu}\phi^{\mu\nu}-\frac{G_{5\tn{X}}}{6}\left[(\square\phi)^3+2\phi_{\mu\nu}^3 -3\phi_{\mu\nu}^2\square\phi\right]\,.
\end{align}
\end{subequations}
Here $g_{\mu\nu}$ is the metric tensor, and $g\equiv {\rm det}(g_{\mu\nu})$ its
determinant. The Ricci scalar and Einstein tensor associated with $g_{\mu\nu}$
are denoted by $R$ and $G_{\mu\nu}$, respectively.
The functions $G_{i}=G_{i}(\phi,X)$ depend only on the scalar
field $\phi$ and its kinetic energy $X=-\partial_\mu\phi\partial^\mu\phi/2$.
We also introduced the shorthand notations
$\phi_{\mu\dots\nu}\equiv \nabla_\mu\dots\nabla_\nu\phi$,
$\phi_{\mu\nu}^2 \equiv \phi_{\mu\nu}\phi^{\mu\nu}$,
$\phi_{\mu\nu}^3 \equiv
\phi_{\mu\nu}\phi^{\nu\alpha}\phi^{\mu}{_{\alpha}}$
and $\Box\phi\equiv g^{\mu\nu} \phi_{\mu\nu}$.

Special cases of Horndeski's theory correspond to well-studied models
of dark energy and modified gravity, including
quintessence~\cite{Ratra:1987rm,Caldwell:1997ii},
k-essence~\cite{ArmendarizPicon:2000ah}, the Dvali-Gabadadze-Porrati
(DGP) model~\cite{Dvali:2000hr,Nicolis:2008in}, and $f(R)$
gravity~\cite{Sotiriou:2008rp,Tsujikawa:2010zza,Clifton:2011jh,Nojiri:2010wj}.
However it is desirable to restrict the large number of functional degrees of
freedom of the action~\eqref{eq:action} by additional theoretical or
phenomenological requirements. For example, it is desirable to
restrict the Horndeski action to models that allow for dynamical
self-tuning of the quantum vacuum energy. This requirement leads to
the Fab Four theory.

\subsection{Fab Four theory}
\label{sec:Fab4}

Starting from the Horndeski action \eqref{eq:action}, Charmousis {\it et
al.}~\cite{Charmousis:2011bf,Charmousis:2011ea} considered homogeneous
isotropic cosmological models satisfying the following requirements:
\begin{enumerate}

\item{The theory admits the Minkowski vacuum for any value of the
    vacuum energy.}

\item{The Minkowski vacuum persists across any phase transition where
    the vacuum energy changes instantaneously by a finite amount.}

\item{The theory admits nontrivial cosmological evolution in the
    presence of matter.}

\end{enumerate}
These requirements lead to the Fab Four action
\begin{equation}
\label{grav}
S = \int d^4 x
\sqrt{-g}
\Big( {\cal L}_\tn{G}[g_{\mu\nu},\phi]+{\cal L}_\tn{M} [g_{\mu\nu},\Psi]\Big),
\end{equation}
where ${\cal L}_\tn{M}[g_{\mu\nu},\Psi]$ is the Lagrangian for matter fields,
collectively represented by $\Psi$, and
\begin{equation}
\label{fab4}
{\cal L}_\tn{G}[g_{\mu\nu},\phi]
=
{\cal L}_{\rm george}
+
{\cal L}_{\rm ringo}
+
{\cal L}_{\rm john}
+
{\cal L}_{\rm paul}\,,
\end{equation}
where
\begin{subequations}
\label{FabFourPot}
\begin{align}
{\cal L}_{\rm george}&=
V_{\rm george}(\phi)R\,,\label{george}\\
{\cal L}_{\rm ringo}&=
V_{\rm ringo}(\phi)R_{\rm GB}\,,\label{ringo}\\
{\cal L}_{\rm john}&=
V_{\rm john}(\phi)G^{\mu\nu}\nabla_\mu\phi\nabla_\nu\phi\,,\label{john}\\
{\cal L}_{\rm paul}&=
V_{\rm paul}(\phi)P^{\mu\nu\alpha\beta}\nabla_\mu\phi\nabla_\alpha\phi
\nabla_\nu\nabla_\beta\phi\,.\label{paul}
\end{align}
\end{subequations}
Here
\begin{eqnarray}
 \label{gb}
 {R}_{\rm GB}\equiv
  R^{\alpha\beta\mu\nu}R_{\alpha\beta\mu\nu}
 -4 R^{\mu\nu}R_{\mu\nu}
 +R^2
\end{eqnarray}
is the Gauss-Bonnet (GB) invariant, and the four potentials
$V_{\rm george} (\phi)$, $V_{\rm ringo} (\phi)$, $V_{\rm john}(\phi)$, and $V_{\rm paul}(\phi)$
are functions of the scalar field.
The quantity
\begin{eqnarray}
\label{dd}
P^{\mu\nu}{}_{\alpha\beta}
&\equiv
-\frac{1}{4}
\delta^{\mu\nu\gamma\delta}_{\sigma\lambda\alpha\beta}
R^{\sigma\lambda}{}_{\gamma\delta}\ ,
\end{eqnarray}
where
\begin{eqnarray}
\label{delta}
\delta^{\mu\alpha\rho\gamma}_{\nu\beta\sigma\delta}
=\left|
\begin{array}{cccc}
\delta^\mu_\alpha &  \delta^\mu_\beta & \delta^\mu_\gamma  & \delta^\mu_\delta \\
\delta^\nu_\alpha  &  \delta^\nu_\beta & \delta^\nu_\gamma   & \delta^\nu_\delta \\
\delta^\rho_\alpha &  \delta^\rho_\beta & \delta^\rho_\gamma & \delta^\rho_\delta \\
\delta^\sigma_\alpha & \delta^\sigma_\beta & \delta^\sigma_\gamma  & \delta^\sigma_\delta  \\
\end{array}
\right|,
\end{eqnarray}
is the double-dual of the Riemann tensor, which shares the symmetries
of the Riemann tensor and satisfies
$\nabla_{\mu}P^{\mu\alpha\nu\beta}=0$.
We assume that $g_{\mu\nu}$ is the Jordan frame metric,
so that the matter fields $\Psi$ do not couple directly to
the scalar field $\phi$.

``George'' reduces to GR and ``Ringo'' -- the
Einstein-dilaton-Gauss-Bonnet (EdGB) term -- becomes trivial in four
dimensions when the respective potentials are constant. Compact
objects in these theories (George and Ringo) have been studied in
detail in the existing literature.  The ``John'' and ``Paul'' terms are more crucial for self-tuning and 
will be the main focus of this paper.

The correspondence between the Horndeski Lagrangians
\eqref{eq:lagrangean} and the Fab Four Lagrangians
\eqref{FabFourPot} was presented in \cite{Charmousis:2011ea}, and we
report it here for completeness:
\begin{subequations}
\label{horndeski_fab4}
\begin{align}
\label{g2ff}
G_2&= 2V_{\rm john}''(\phi)X^2-V_{\rm Paul}^{(3)} (\phi) X^3
+6 V_{\rm george}''(\phi) X
\nonumber \\
&+8 V_{\rm ringo}^{(3)} (\phi)X^2 \big(3-\ln (|X|)\big),
\\
\label{g3ff}
G_3&= 3V_{\rm john}'(\phi)X
-\frac{5}{2}V_{\rm paul}''(\phi)X^2
+3 V_{\rm george}'(\phi)
\nonumber\\
&+4 V_{\rm ringo}^{(3)} (\phi)
  X \big(7-3\ln (|X|)\big),
\\
\label{g4ff}
G_4&= V_{\rm john}(\phi) X
-V_{\rm paul}'(\phi)X^2
+V_{\rm george} (\phi)
\nonumber\\
&+4 V_{\rm ringo}''(\phi)
  X \big(2-\ln (|X|)\big),
\\
\label{g5ff}
G_5&=
-3 V_{\rm paul}(\phi) X
-4 V_{\rm ringo}'(\phi)\ln (|X|).
\end{align}
\end{subequations}

\subsection{Cosmology and Bback holes in Fab Four theory}

Cosmological evolution in Fab Four gravity in the presence of ordinary
matter and radiation has been exhaustively investigated by Copeland 
{\it et al.}~\cite{Copeland:2012qf}.  They demonstrated that for a specific
choice of the Fab Four potentials in Eq.~\eqref{FabFourPot}, even if
the source is dominated by the vacuum energy and there is no explicit
matter fluid, the cosmological evolution toward the self-tuned
Minkowski attractor can mimic the matter-dominated evolution
corresponding to
dark matter.
Moreover, Refs.~\cite{Gubitosi:2011sg,Appleby:2012rx} demonstrated the
existence of a self-tuned de Sitter (dS) attractor for a certain
nonlinear combination of the canonical kinetic term to the Fab Four.
References~\cite{Martin-Moruno:2015bda,Martin-Moruno:2015lha} presented a
systematic derivation of the most general subclass of Horndeski's
theory that can allow for a spatially flat self-tuned dS vacuum.  This
new subclass of Horndeski's theory is expected to have a deep
connection to the Fab Four theory, but it was derived in an independent
way and their relation remains unclear.
A specific form of John and
Paul also appears in a proxy theory to nonlinear massive gravity
\cite{deRham:2011by}, but a close inspection of cosmological dynamics
revealed that there is no de Sitter attractor in this model~\cite{Heisenberg:2014kea}.

A challenge to the Fab Four model is that self-tuning has been
verified only for homogeneous, isotropic cosmological backgrounds.
The presence of stars and black holes (BHs) in the Universe implies
that self-tuning should still occur in the presence of local
inhomogeneities of the spacetime, such as point masses or extended
self-gravitating bodies. Whether self-tuning occurs in
inhomogeneous spacetimes is a nontrivial question.

A first step towards answering this question is the investigation of
BH solutions in Fab Four theory. Most studies of BH solutions in
Horndeski's theory and Fab Four gravity have focused on the
shift-symmetric subclass of the theories. An influential work by Hui
and Nicolis~\cite{Hui:2012qt} proved a BH no-hair theorem in
Horndeski gravity.
The theorem makes the following assumptions:
(i) the spacetime  is static and  spherically symmetric;
(ii) the scalar field shares the same symmetries as the spacetime,
i.e. $\phi=\phi(r)$, where $r$ is the radial coordinate;
(iii) the theory is shift-symmetric (i.e. it is invariant under the transformation
$\phi\to \phi+c$, where $c$ is a constant);
and (iv) the spacetime is asymptotically flat.

Searches for hairy BH solutions followed two main routes: they either
looked for loopholes in the Hui-Nicolis theorem, or relaxed the
assumptions behind the theorem.  All BH solutions found so far in
Horndeski's theory have secondary hair, i.e. the scalar charge is not
independent of other charges, such as the mass (see
e.g.~\cite{Herdeiro:2015waa} for a review of BH solutions with scalar
hair).

Sotiriou and Zhou found a loophole in the Hui-Nicolis no-hair
theorem~\cite{Sotiriou:2013qea,Sotiriou:2014pfa}.  In our language,
they considered the combination George$+$Ringo with
$V_{\rm george}={\rm constant}$ and $V_{\rm ringo}\propto \phi$ in
Eq.~\eqref{ringo}.  Other authors relaxed assumption (iv), finding
asymptotically anti-de Sitter (AdS) BH solutions for actions of the
John type (nonminimal coupling to the Einstein tensor) with
$V_{\rm john}={\rm
  constant}$~\cite{Rinaldi:2012vy,Anabalon:2013oea,Minamitsuji:2013ura}
(see~\cite{Minamitsuji:2014hha,Cisterna:2015uya} for a stability
analysis of BH solutions in theories of the John subclass).  BH
solutions that may be more relevant for astrophysics were found by
Babichev and Charmousis~\cite{Babichev:2013cya} for theories of the
George$+$John type, with $V_{\rm george}$ and $V_{\rm john}$ both
constant, relaxing assumption (ii).
Babichev and Charmousis introduced a linear time dependence in the
scalar field that therefore does not possess the same symmetries as
the metric. However the effective energy-momentum tensor remains
static because of the shift symmetry.  A particularly important
asymptotically flat BH solution emerging from this analysis is a
``stealth'' solution in the George$+$John class: a Schwarzschild
BH metric supports a nontrivial, regular scalar field configuration
which does not backreact on the spacetime.  By adding the canonical
kinetic term for the scalar field and the cosmological constant
$\Lambda$, Babichev and Charmousis also obtained a Schwarzschild-(A)dS
solution.  Interestingly, the effective cosmological constant one can
read off from the Schwarzschild-(A)dS metric does {\em not} depend on
$\Lambda$, and the $\Lambda$ dependence appears only in the scalar
field. Therefore this solution may be interpreted as an extension of
the self-tuned dS vacuum to an inhomogeneous spacetime.

In Ref.~\cite{Maselli:2015yva}, all of the above static, spherically
symmetric BH solutions were generalized to slow rotation at leading
order in the Hartle-Thorne approximation~\cite{Hartle:1967he,Hartle:1968si}.
For all of these solutions,
first-order corrections due to rotation were shown to be identical to
GR. The Hui-Nicolis no-hair theorem was extended to slowly rotating
BHs for which the scalar field is allowed to have a linear time
dependence.  Moreover, all the spherically symmetric solutions
obtained for the John class can be naturally extended to the
shift- and reflection-symmetric subclass of Horndeski's theory, namely
theories with $G_2=G_2(X)$, $G_4=G_4(X)$, and
$G_3=G_5=0$~\cite{Kobayashi:2014eva}.

In summary, nontrivial BH solutions in Fab Four gravity were found for
the Ringo and John subclasses. In particular, the Schwarzschild-dS
solution found in the case of nonminimal coupling with the Einstein
tensor (John) can be seen as a self-tuned BH solution.  On the other
hand, to our knowledge, no analytic or numerical BH solutions have
been reported for the Paul subclass.  Because of the similarity
between John and Paul, one may naively expect that Paul should also
allow for self-tuned, inhomogeneous vacuum solutions.  This question
was partially addressed by Appleby~\cite{Appleby:2015ysa}, who claimed
that self-tuned BH solutions would not exist in the Paul case.  This
is because in a Schwarzschild-dS spacetime the Weyl components of
$P^{\mu\nu}{}_{\alpha\beta}$ and $R_{\rm GB}$ terms in the scalar
field equation of motion contain an explicit dependence on the radial
coordinate, and leave no redundancy in the scalar field equation of
motion.  This is in contrast to the case of ``John,'' where the scalar
field equation of motion contains no Weyl component that could make
it redundant for a Schwarzschild-dS metric. This also hints at the
absence of similar BH solutions in the non-reflection-symmetric
subclass of the shift-symmetric Horndeski theory with nonzero $G_3(X)$
and $G_5(X)$, although there are no detailed studies of this issue.

\subsection{Plan of the paper}

The next natural step to test whether the Fab Four model is compatible
with local inhomogeneities is to consider self-gravitating matter
configurations, and in particular static or rotating neutron stars
(NSs). The main goal of this paper is precisely to investigate the
existence and properties of slowly rotating NS solutions in Fab Four
gravity.

The structure and stability of rotating NSs in GR (George) is, of
course, textbook
material~\cite{Shapiro:1983du,FriedmanStergioulas,Stergioulas:2003yp}.
In the past few years there has been significant progress in our
understanding of slowly ~\cite{Pani:2011xm} and rapidly
rotating~\cite{Kleihaus:2014lba,Kleihaus:2016dui} NSs in
Einstein-dilaton-Gauss-Bonnet gravity (Ringo), and there are also
studies of stellar stability under odd-parity (axial) perturbations in
this theory~\cite{Blazquez-Salcedo:2015ets}.  Recent investigations
turned to theories with nonminimal coupling to the Einstein tensor
(John)~\cite{Cisterna:2015yla,Silva:2016smx,Cisterna:2016vdx}. Here we
complete and extend the analysis of NSs in the John subclass, and we
look for solutions in theories containing the Paul term.
We were unable to obtain NS solutions in theories involving the Paul term.
Apparently, Paul does not want to be a star.


This paper is organized as follows. In Sec.~\ref{sec:EOM} we derive
the stellar structure equations at first order in a slow-rotation
expansion in generic shift-symmetric Horndeski theories. In
Sec.~\ref{sec:fabfournss} we specialize our analysis to each of the
Fab Four subcases.
In Sec.~\ref{sec:conclusions} we summarize our findings and point
out possible directions for future research. Appendix~\ref{app:IC}
discusses the relation between the moment of inertia and the stellar
compactness in theories of the Ringo and John subclasses. Throughout
the paper, unless specified otherwise, we will use geometrical units
($G=c=1$).


\section{Slowly rotating stars in Fab Four theory}
\label{sec:EOM}

In this section we will consider the shift-symmetric subclass of
Horndeski's theory that is invariant under the transformation
\begin{equation}
\phi\rightarrow \phi + c\,,
\end{equation}
where $c$ is a constant.  From Eqs.~\eqref{horndeski_fab4}, this
assumption implies that $V_{\rm john}$, $V_{\rm paul}$, and
$V_{\rm george}$ must be constant, while the Ringo (EdGB) term
$V_{\rm ringo}$ can be a linear function of $\phi$. For EdGB, a
constant shift in $\phi$ only adds a trivial topological invariant to
the action, and therefore it does not affect the field equations.
Equations~\eqref{fab4} and \eqref{FabFourPot} represent the basic building
blocks of our theory, which will be described by the general action
\begin{equation}
{S} = {S}_\tn{G}+{S}_\tn{M}\,,\label{genaction}
\end{equation}
where ${S}_\tn{M}$ is the ordinary action for fluid matter and
${S}_\tn{G}$ is a combination of the Lagrangians
\eqref{john}-\eqref{ringo}.

To investigate slowly rotating solutions we follow the approach
described by Hartle and Thorne~\cite{Hartle:1967he,Hartle:1968si}, in which spin
corrections are considered as small perturbations on an otherwise static,
spherically symmetric background. In particular, at first order in the
star's angular velocity $\Omega$ the metric can be written as
\begin{align}
ds^2&=-A(r)dt^2+\frac{dr^2}{B(r)}+r^2d\theta^2+r^2\sin^2\theta d\varphi^2\nonumber\\
&-2[\Omega-\tilde{\omega}(r)]\sin^2\theta dtd\varphi\,,\label{metric}
\end{align}
where $\tilde{\omega}(r)$ is the angular velocity of the fluid as measured
by a freely falling observer.

Varying the action \eqref{genaction} with respect to the metric and
the scalar field we obtain the equations of motion for
$g_{\alpha\beta}$ and $\phi$, respectively:
\begin{equation}
\mathcal{E}_{\alpha\beta}=T_{\alpha\beta}\,,
 \quad \mathcal{E}_{\phi}=0\,,\label{EoM}
\end{equation}
where
\begin{equation}
T_{\alpha\beta}=(\epsilon+p)u_\alpha u_\beta+p g_{\alpha\beta}
\end{equation}
is the energy-momentum tensor of a perfect fluid. Here $\epsilon$ and
$p$ are the energy density and pressure of a fluid element with
four-velocity $u^\mu=u^0(1,0,0,\Omega)$. The time component $u^0$
follows directly from the normalization condition $u^\mu u_\mu=-1$,
which leads for the metric \eqref{metric} to $u^0=1/\sqrt{A}$.  The
explicit form of $\mathcal{E}_{\alpha\beta}$ and $\mathcal{E}_{\phi}$
can be found in the Appendix of \cite{Maselli:2015yva}
(see \cite{Cisterna:2015uya} for a particular study in the case of John).

Moreover, in the Jordan frame, the energy-momentum tensor is conserved:
\begin{equation}
\nabla_\mu T^{\mu\nu}=0 \,.\label{enecons}
\end{equation}
To close the system of equations we need to specify the equation of
state (EOS) for the NS, i.e., a relation between the pressure and
energy density:
\begin{equation}
p=p(\epsilon)\,.\label{EOS}
\end{equation}
Taken together, Eqs.~\eqref{EoM}, \eqref{enecons}, and \eqref{EOS}
provide the full description of a slowly rotating star.

In this work we will consider three realistic EOSs, namely, APR
\cite{Akmal:1998cf}, SLy4 \cite{Douchin:2001sv} and GNH3 \cite{Glendenning:1984jr}
in decreasing order of stiffness. To facilitate comparisons with
\cite{Cisterna:2015yla,Cisterna:2016vdx} we will also consider a
polytropic EOS of the form $p=K\rho^\Gamma$, with $K=123M_\odot^2$ and
$\Gamma=2$. Here $\rho$ is the mass density, related to the energy density
by
\begin{equation}
\epsilon=\left(\frac{p}{K}\right)^{1/\Gamma}+\frac{p}{\Gamma-1}\,.
\end{equation}
In Table~\ref{tableEOS} we show the radius $R$ and compactness
${\cal C}\equiv M/R$ of nonrotating models, as well as the moment of inertia
$I$, for NSs with the ``canonical'' mass $M=1.4\,M_\odot$ constructed
using different EOS models in GR.  At fixed mass, the realistic EOSs
APR, SLy4, and GNH3 (in this order) yield
configurations with decreasing compactness, and therefore larger
moment of inertia.
\begin{table}[ht]
\centering
\begin{tabular}{cccc}
\hline
\hline
EOS &
$R$ (km)
& ${\cal C}$
& $I$ ($10^{45}$g cm $^2$) 
\\
\hline
APR &
11.33 &
0.182 &
1.31 
\\
SLy4
&11.72
&0.176
&1.37
\\
GNH3
& 14.18
&0.146
& 1.81 
\\
Polytrope
& 16.48
& 0.125
& 2.28
\\
\hline
\hline
\end{tabular}
\caption{The radius $R$, compactness ${\cal C}$, and
  moment of inertia $I$ for a canonical NS with mass $M=1.4\,M_\odot$, in GR,
  using three different nuclear-physics based
  EOS models and a $\Gamma=2$ polytrope.}
\label{tableEOS}
\end{table}

\section{Fab Four neutron stars}
\label{sec:fabfournss}

In this section we discuss NSs in the four subclasses of Fab Four
gravity, starting from the simplest Lagrangians.

\subsection{George\\(General relativity)}
\label{sec:GR}

The George Lagrangian for shift-symmetric theories corresponds to
GR, so we refer the reader to standard treatments of rotating
stars~\cite{Shapiro:1983du,FriedmanStergioulas,Stergioulas:2003yp}.


\subsection{Ringo\\(Einstein-dilaton-Gauss-Bonnet gravity)}

Nonrotating hairy BH solutions in EdGB gravity with a dilatonic
coupling of the schematic form
$V_{\rm ringo}\sim \zeta e^{\gamma \phi}$ were found by Kanti {\it et
al.}~\cite{Kanti:1995vq}. These solutions were then extended to slowly
and rapidly rotating BHs~\cite{Pani:2011xm,Kleihaus:2011tg}.
As stated in the introduction, Sotiriou and
Zhou~\cite{Sotiriou:2013qea,Sotiriou:2014pfa} pointed out that hairy
BH solutions exist in shift-symmetric EdGB theories, in violation of
the Hui-Nicolis no-hair theorem (see~\cite{Maselli:2015yva} for an
extension of these results to linear order in a slow-rotation
approximation). Shift-symmetric EdGB theories can be seen as a
small-field Taylor series expansion of the dilatonic coupling
\begin{equation}
\label{EdGBTaylor}
V_{\rm ringo}\simeq \zeta+\zeta\gamma\phi\,,
\end{equation}
where the constant term $\zeta$ can be neglected since it gives rise
to a topological invariant at the level of the action.

NSs in EdGB gravity with a dilatonic coupling were studied
in~\cite{Pani:2011xm,Kleihaus:2014lba,Kleihaus:2016dui} (see
also~\cite{Blazquez-Salcedo:2015ets} for axial perturbations).  As it
turns out, the bulk properties of NSs depend only on the combination
$\zeta\gamma$; cf. the discussion around Eq.~(29) of
\cite{Pani:2011xm}.  This is because the value of the scalar field is
typically very small within the star, and therefore the Taylor
expansion \eqref{EdGBTaylor} is an excellent approximation. For this
reason, the analysis of NSs in Ref.~\cite{Pani:2011xm} applies also to
the shift-symmetric case of interest here, and we refer the reader to
the treatment in that paper for calculations of stellar structure and
observational bounds on the product $\zeta\gamma$.

\label{sec:edgb}

\subsection{John\\(Nonminimal coupling with the Einstein tensor)}
\label{sec:nonmin}

A more interesting case is slowly rotating compact stars in theories
with a nonminimal derivative coupling with the Einstein tensor,
corresponding to the John Lagrangian
\eqref{john}~\cite{Cisterna:2015yla,Silva:2016smx,Cisterna:2016vdx}.
These theories are described by the action
\begin{align}
{S_\tn{G}}&=\int d^4x \sqrt{-g} ({\cal L}_{\rm george}+{\cal L}_{\rm john}+{\cal L}_\tn{K})\nonumber\\
&=\int d^4x\sqrt{-g}
\left[\kappa R-\frac{1}{2}(\beta g^{\mu\nu}-\eta G^{\mu\nu})\partial_\mu\phi\partial_\nu\phi\right],\nonumber \\ \label{nonminact}
\end{align}
where
${\cal L}_\tn{K}=
\beta X=-(
\beta\partial_\mu\phi\partial^\mu\phi)/2$
is a kinetic term for the scalar field, $\beta$ and $\eta$ are
constants, and $\kappa=(16\pi)^{-1}$.
Equation~\eqref{nonminact} can be
obtained from the Horndeski Lagrangian by choosing
\begin{equation}
\label{g2g4john}
G_2=\beta X ,\quad G_4=\kappa+\frac{\eta}{2}X\ , \quad
G_3=G_5=0\ .
\end{equation}
We also consider a real scalar field of the form~\cite{Babichev:2013cya}
\begin{equation}\label{ansatzphi}
\phi(r,t)=q t+\psi(r) \ ,
\end{equation}
where $q$ is a constant scalar charge. With this choice, the field's
kinetic energy is a function of $r$ only:
\begin{equation}
X=
\frac{1}{2}\left[\frac{q^2}{A(r)}-B(r)\psi'(r)^2\right]\ .
\end{equation}

In vacuum, the theory described by the action \eqref{nonminact} leads
to asymptotically AdS black hole solutions with a
nontrivial scalar field configuration
\cite{Rinaldi:2012vy,Minamitsuji:2013ura,Anabalon:2013oea,Kobayashi:2014eva,Babichev:2013cya}.
However, it has recently been shown that it is possible to construct
``stealth'' NS models for which the exterior solution is given by the
Schwarzschild spacetime \cite{Cisterna:2015yla}.

For $\beta=0$, the scalar field outside the star (where
$T_{\mu\nu}=0$) does not backreact on the metric, leading to ``stealth
solutions''. However inside the star (where $T_{\mu\nu}\neq0)$ the
scalar field has a nontrivial effect, and the stellar structure is
different from GR.

Hereafter we will focus on these stealth solutions, fixing
$\beta=0$. We recall that the action \eqref{nonminact} is invariant
under shift symmetry ($\phi\rightarrow \phi+c$). This allows us to
write the equation of motion for the scalar field in terms of a
conserved current $J^{\mu}$:
\begin{equation}\label{eq1}
\nabla_\mu J^\mu=0\,,
\end{equation}
with nonzero components given by
\begin{align}
J^t=&-\frac{q\eta}{r^2\kappa A}(rB'+B-1)\,,\\
J^r=&\frac{\eta B}{r^2\kappa A}[A(B-1)+rBA']\phi'\,.\label{eqJr}
\end{align}
We also remark that Eq.~\eqref{eq1}, using the line element
\eqref{metric}, admits the solution
\begin{equation}
J^r=\sqrt{\frac{B}{A}}\frac{C_1}{r^2}\,,
\end{equation}
with $C_1$ constant. In the following we will set $C_1=0$, as it has
been shown that this choice is consistent with a vanishing radial
energy flux, i.e., ${\cal E}_{tr}=0$ \cite{Babichev:2015rva}.

Combining Eqs.~\eqref{enecons}, \eqref{eqJr}, and the $(tt)$ and $(rr)$
components of Eqs.~\eqref{EoM}, we obtain a set of differential
equations for the spherically symmetric background.  Moreover, at
linear order in the angular velocity, the $(t\varphi)$ equation
$\mathcal{E}_{t\varphi}-T_{t\varphi}=0$ yields a differential equation
for $\tilde{\omega}$. In summary, a slowly rotating NS at first order
in the slow-rotation approximation is described by the following set
of equations:
\begin{align}
A'=&\frac{A}{r}\frac{1-B}{B}\ ,\label{ODE1}\\
B'=&\frac{3q^2\eta B(B-1)-A[r^2 \epsilon-4\kappa+B(4\kappa+r^2(\epsilon+6p))]}{r[A(4\kappa+r^2p)-3q^2\eta B]}\ ,\label{ODE2}\\
p'=&-\frac{\epsilon+p}{2}\frac{A'}{A}\ ,\label{ODE3}\\
\tilde{\omega}''=&\frac{4q^2\eta B^2-A[4B(4\kappa+r^2p)-r^2(\epsilon+p)]}{rB[A(4\kappa+r^2p)-q^2\eta B]}\tilde{\omega}'\nonumber\\
&-\frac{4A(\epsilon+p)}{B[q^2\eta B-A(4\kappa+r^2p)]}\tilde{\omega}\ ,\label{ODE4}\\
(\phi')^2&=\frac{r^2Ap-q^2\eta(B-1)}{\eta A B}\ .\label{ODE5}
\end{align}
Note that $q$ and $\eta$ always appear combined in the factor $q^2 \eta$.

\begin{figure*}[th]
\includegraphics[width=\columnwidth]{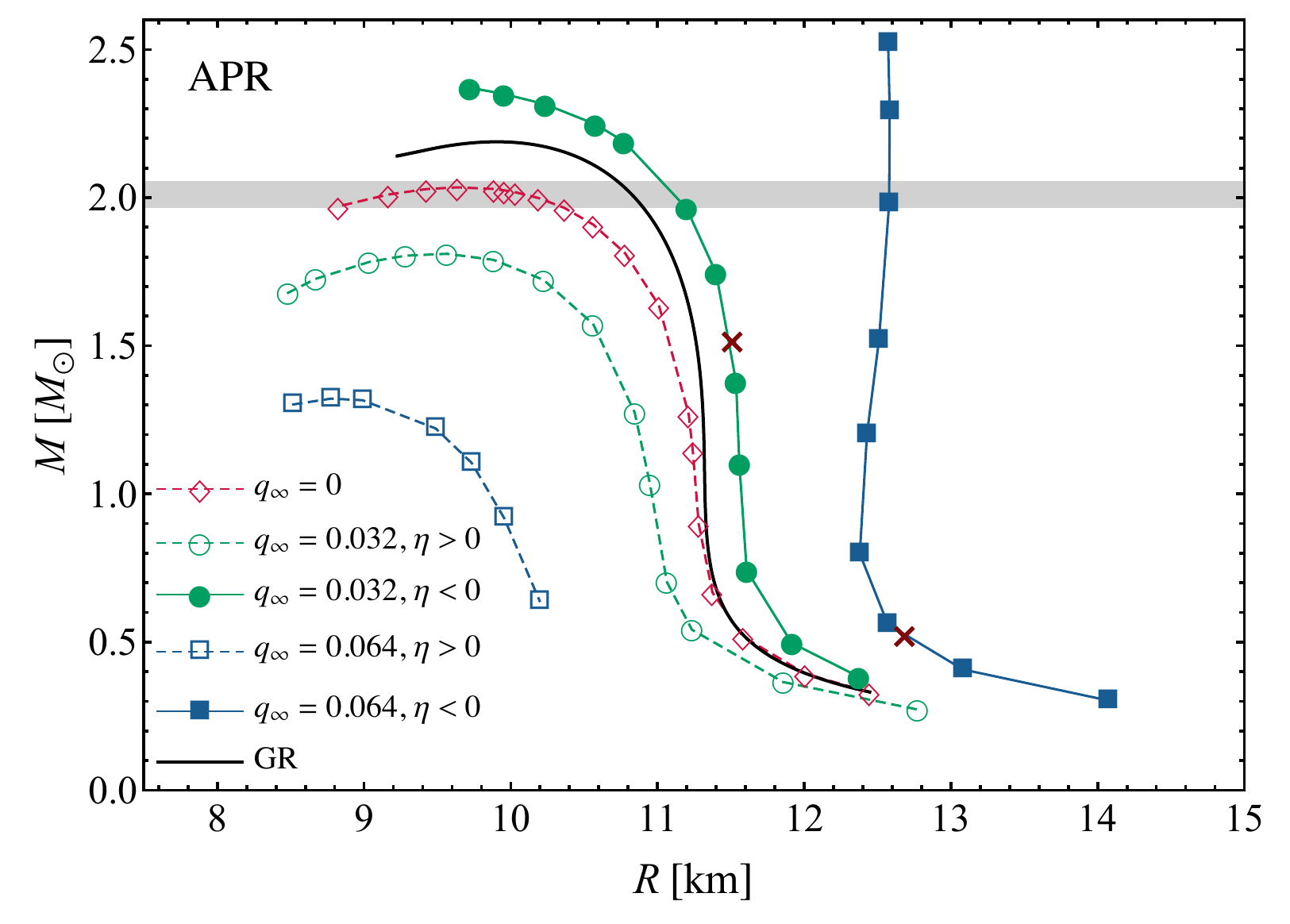}
\includegraphics[width=\columnwidth]{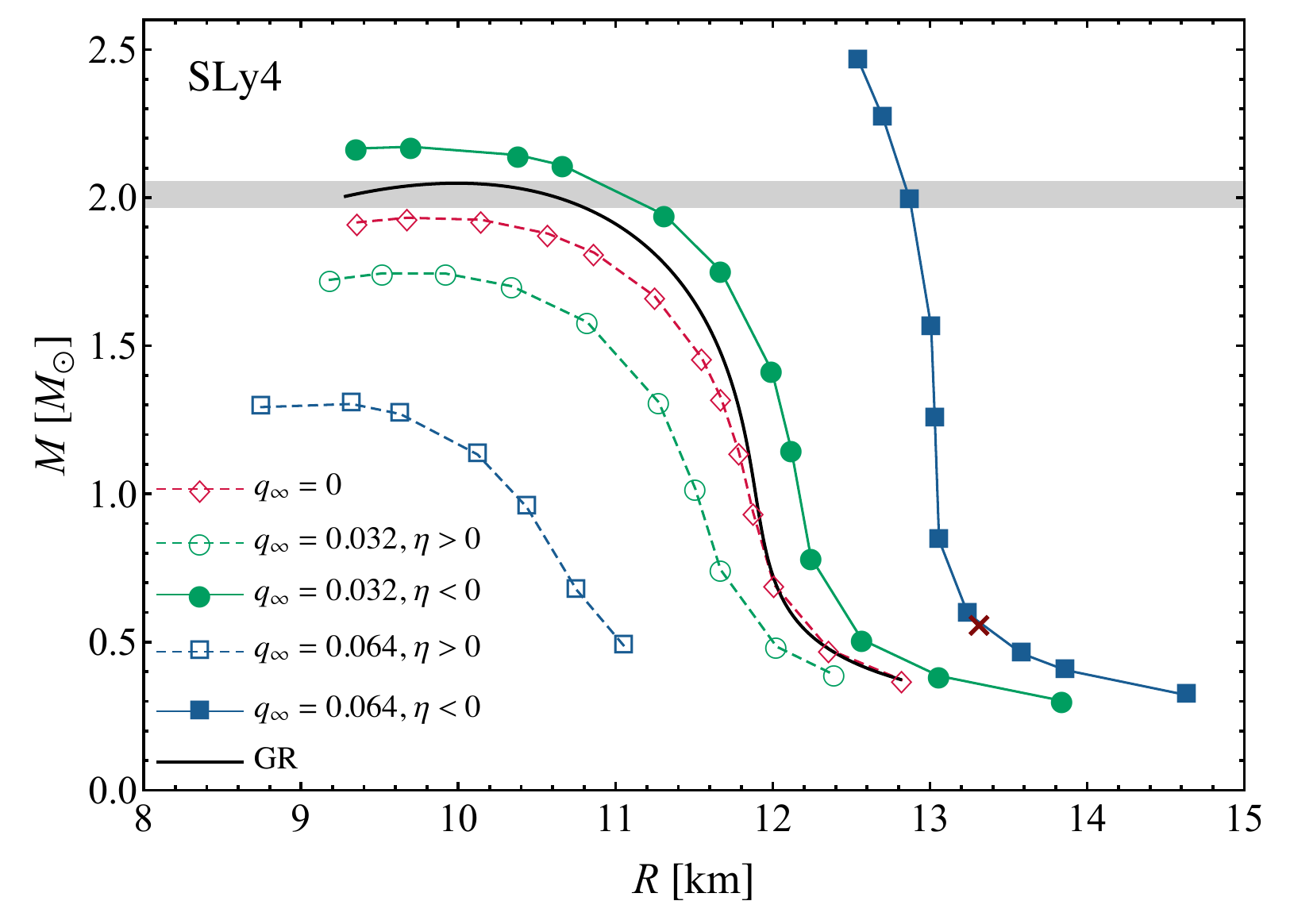} \\
\includegraphics[width=\columnwidth]{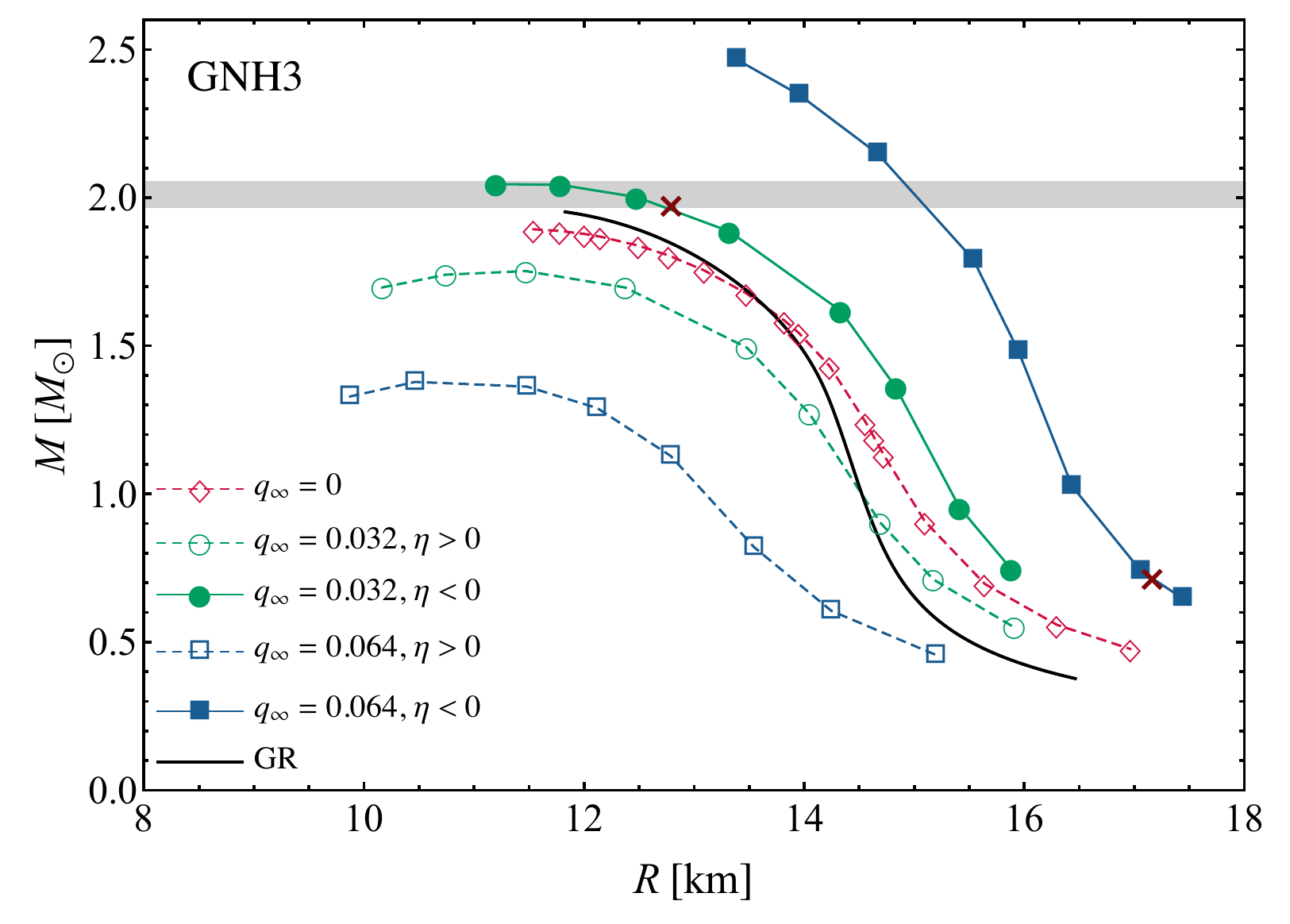}
\includegraphics[width=\columnwidth]{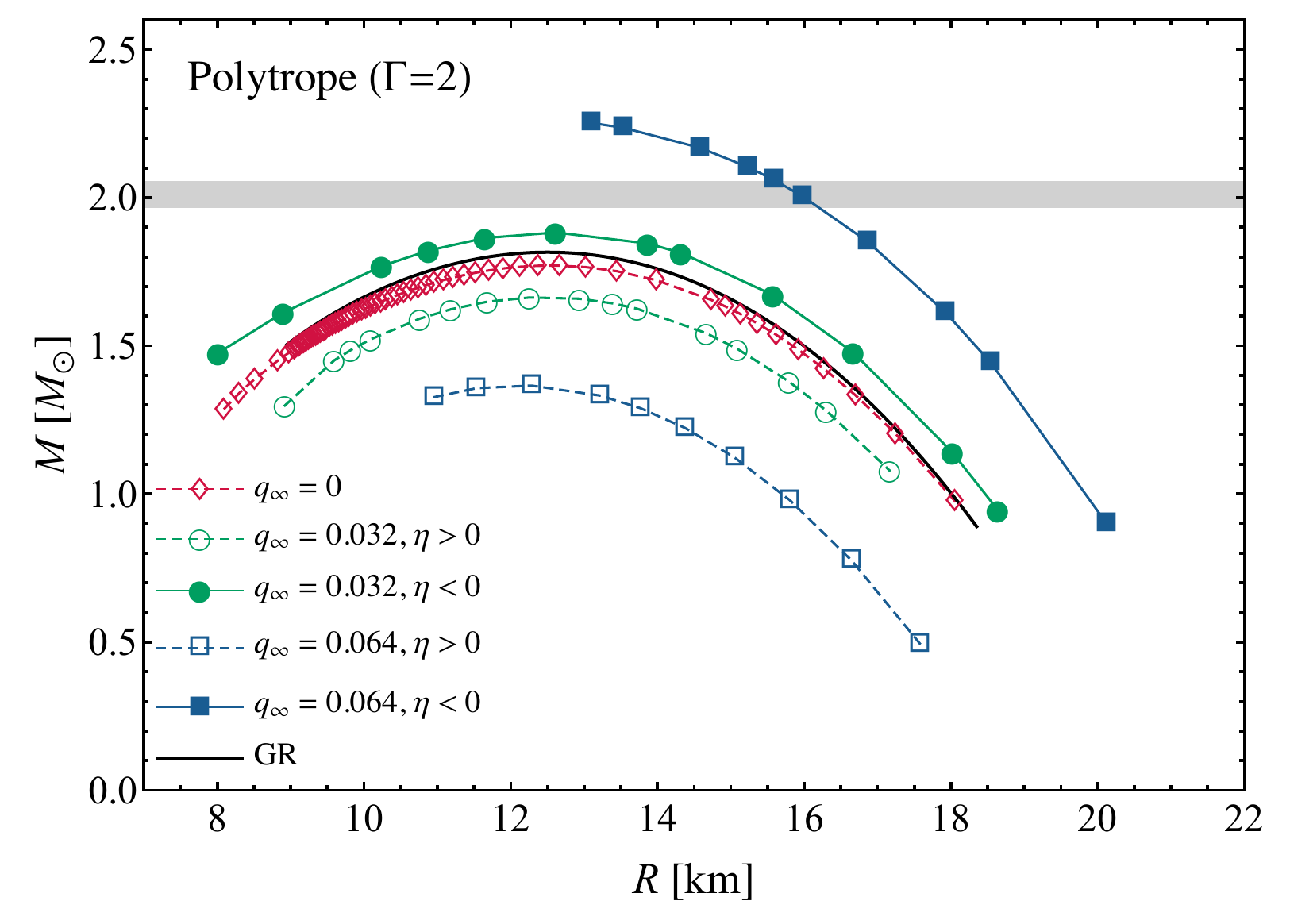}
\caption{Mass-radius curves for different EOS models, selected values
  of $q_\infty$, and $\eta=\pm 1$.  The various panels correspond to
  EOS APR (top left), SLy4 (top right), GNH3 (bottom left) and a
  polytropic (bottom right).  Configurations with radii smaller than
  that identified by the orange cross do not satisfy the condition
  \eqref{eq:cond2}.  The horizontal colored band corresponds to
  $M = 2.01 \pm 0.04$ $M_{\odot}$, the most massive NS mass known to
  date~\cite{Demorest:2010bx}.  Note that the various panels have
  different $x$-axis ranges.}
\label{fig:massradius}
\end{figure*}

\begin{figure*}[th]
\includegraphics[width=\columnwidth]{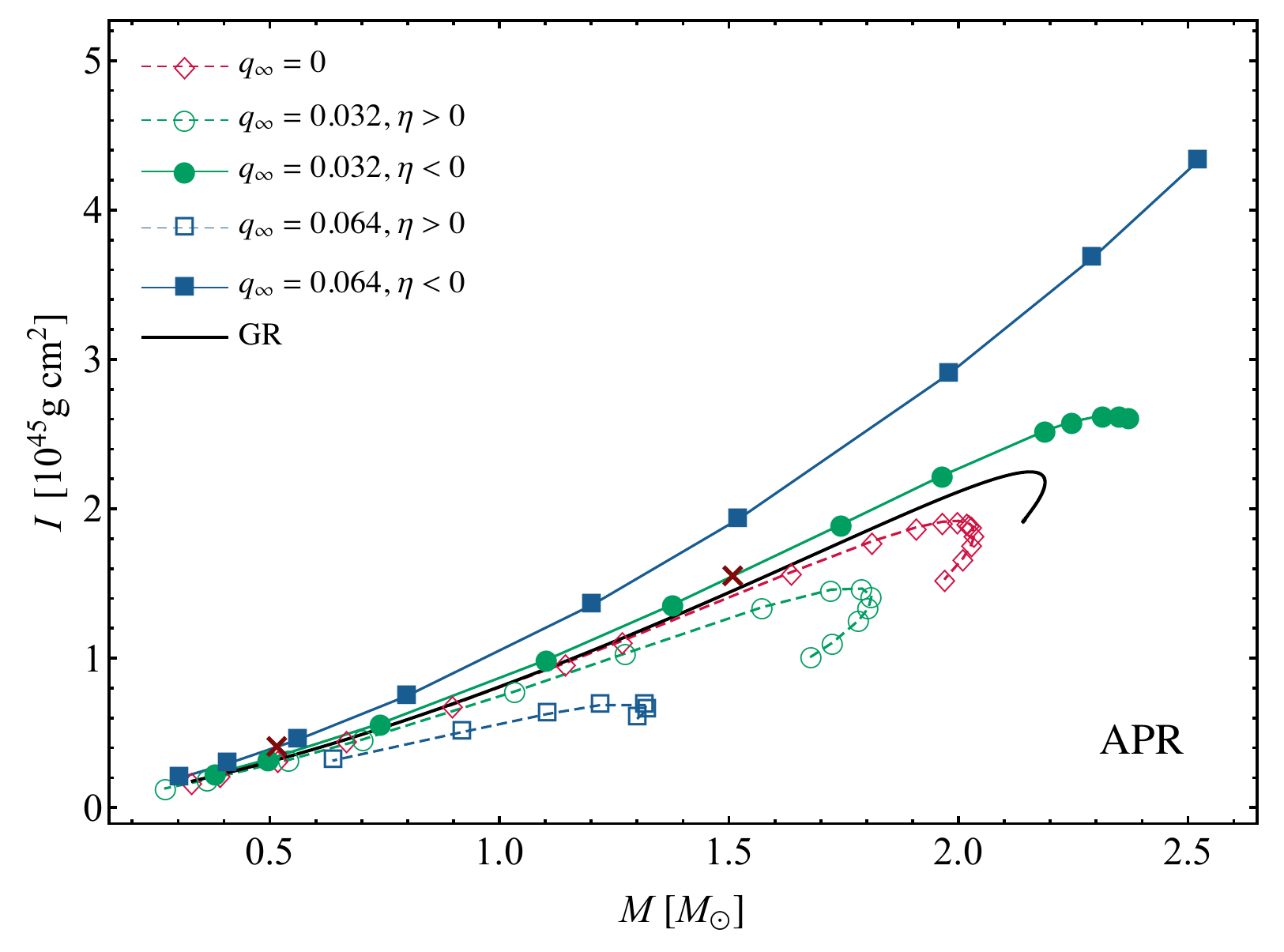}
\includegraphics[width=\columnwidth]{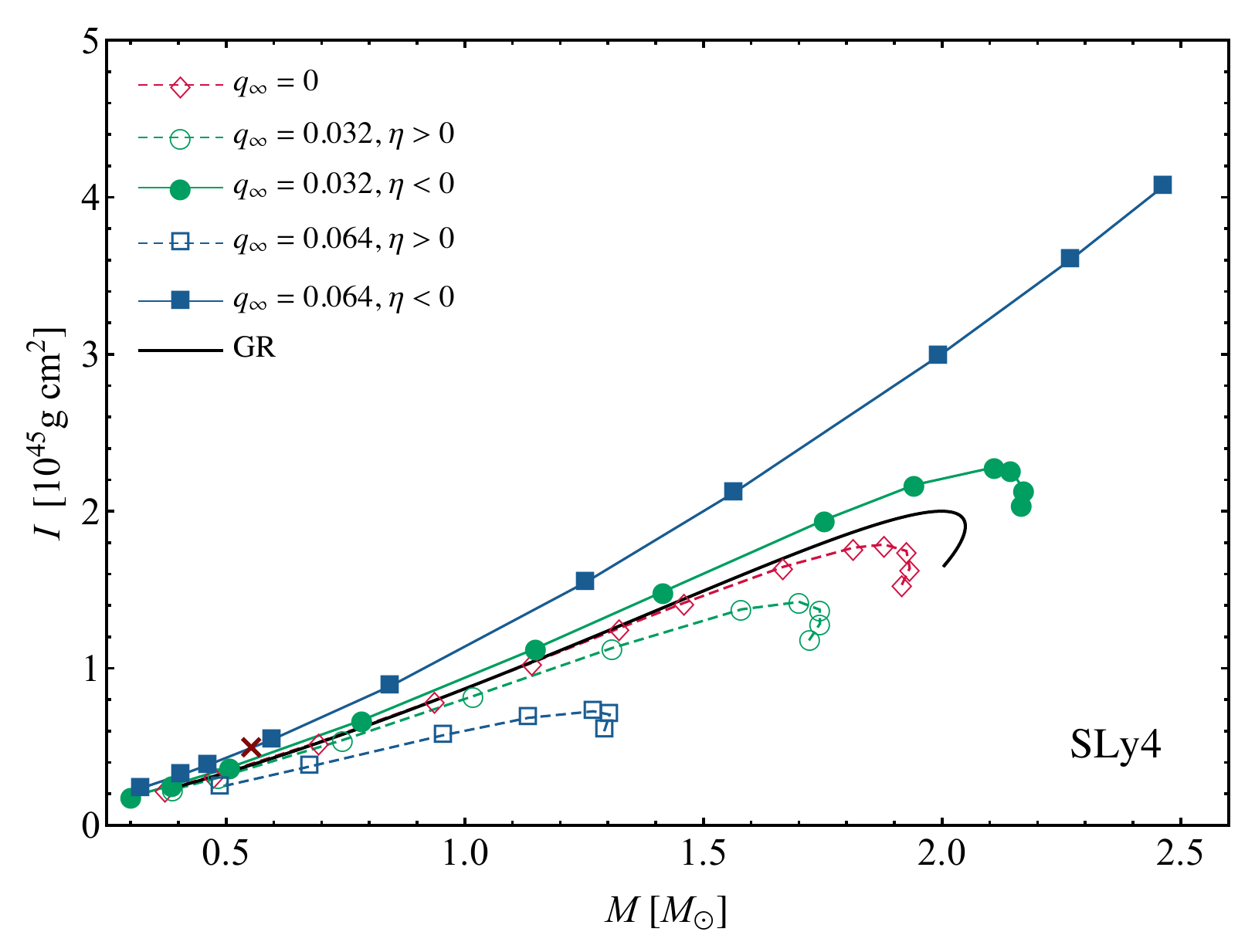}\\
\includegraphics[width=\columnwidth]{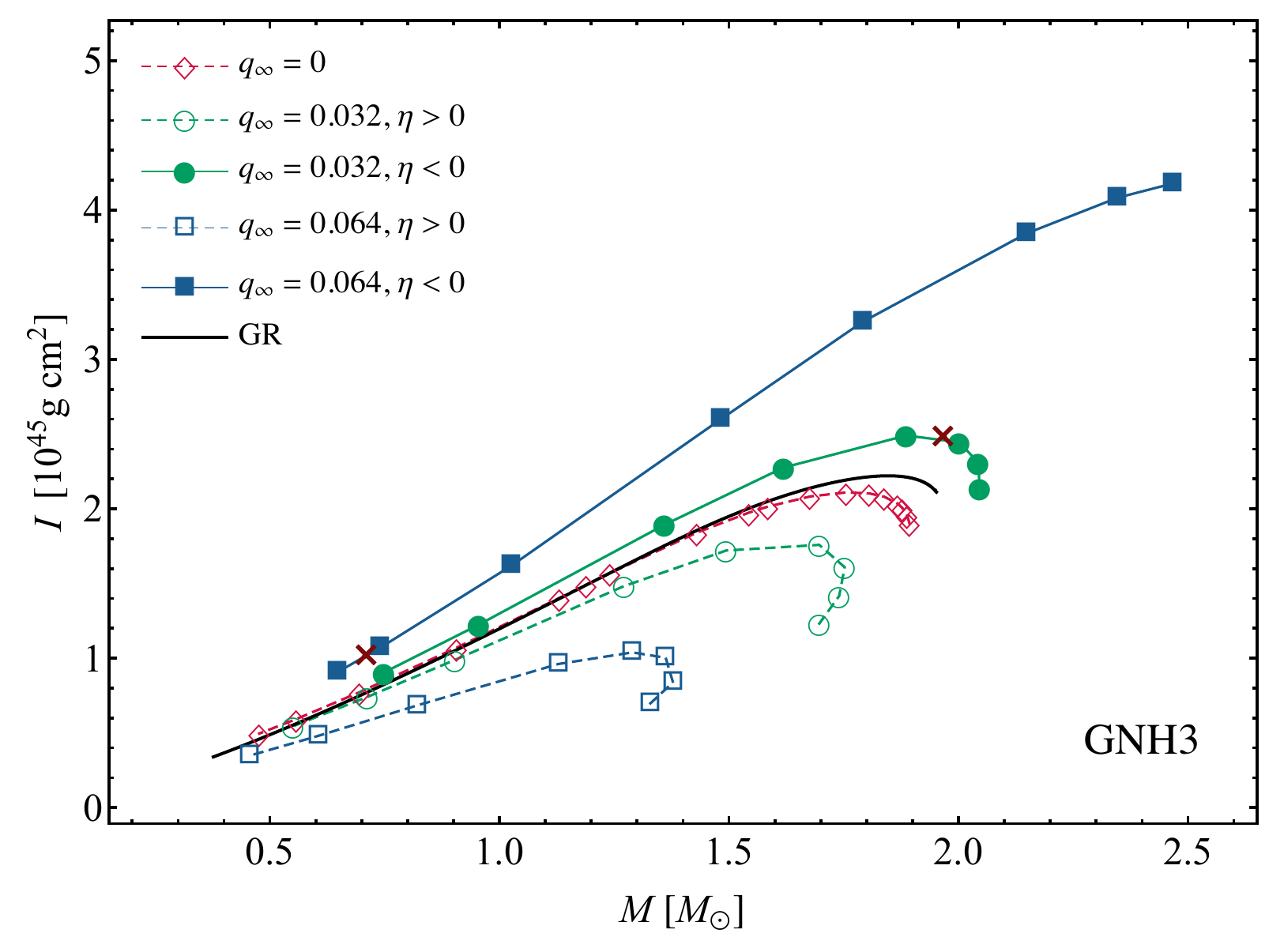}
\includegraphics[width=\columnwidth]{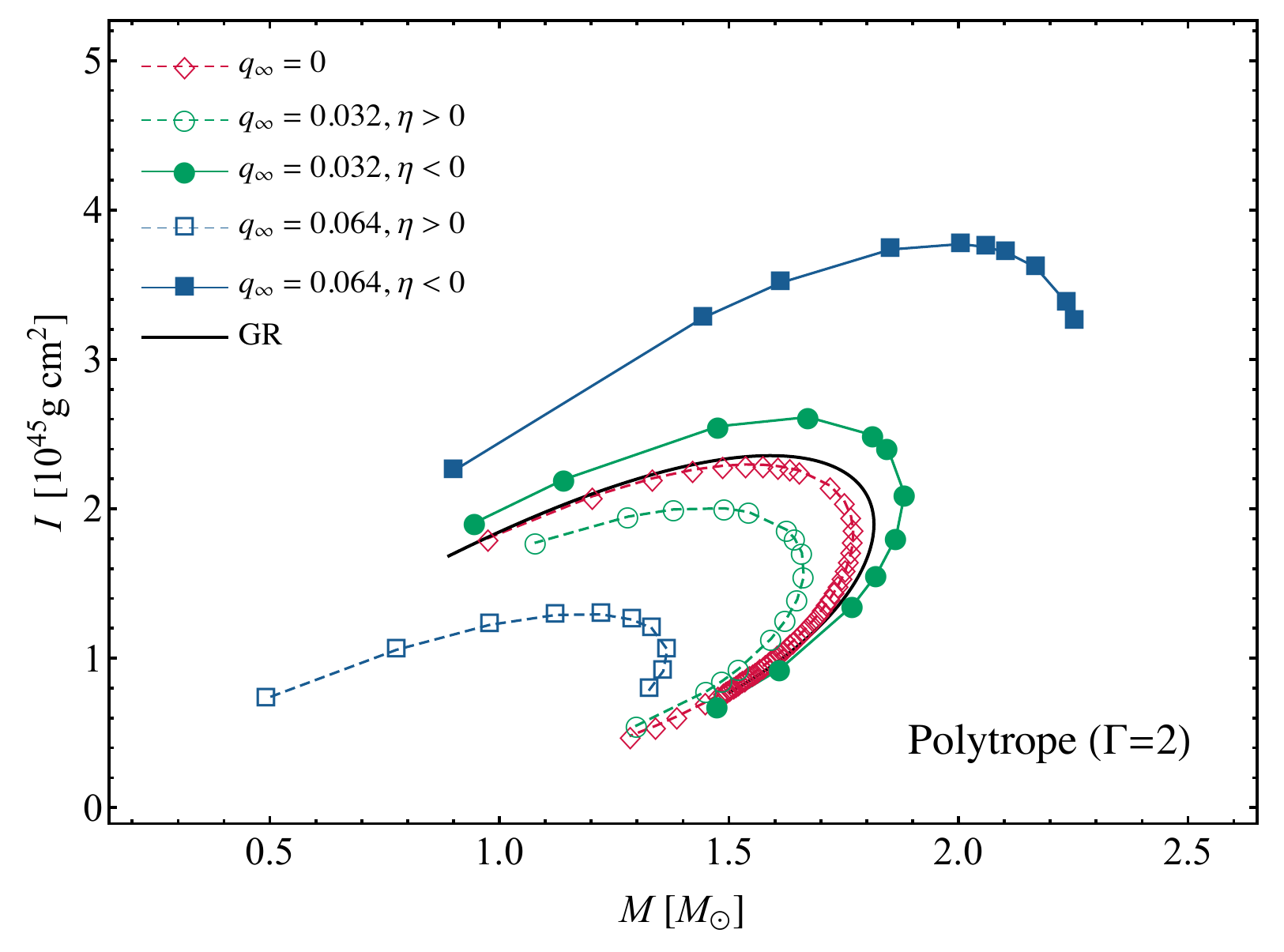}
\caption{Moment of inertia $I$ as a function of the mass $M$.  The
  various panels correspond to EOS APR (top-left), SLy4 (top-right),
  GNH3 (bottom left) and a polytropic (bottom right).  Configurations
  with masses larger than that identified by the orange cross do not
  satisfy the condition \eqref{eq:cond2}.}
\label{fig:inertiamass}
\end{figure*}

Expanding all variables in a power series around $r=0$, we obtain the initial
values for $(A,B,\tilde{\omega},p,\phi)$ as
\begin{subequations}
\begin{align}
A&=A_c-\frac{r^2A_c^2(3p_c+\epsilon_c)}{3(3q^2\eta-4\kappa A_c)} + \mathcal{O}(r^3)\,,\label{B1}\\
B&=
1+\frac{2}{3}\frac{r^2A_c(3 p_c+\epsilon_c)}{(3q^2\eta-4\kappa A_c)}+\mathcal{O}(r^3)\,,\label{B2}\\
p&=p_c+\frac{r^2A_c(p_c+\epsilon_c)(3p_c+\epsilon_c)}{6(3q^2\eta-4\kappa A_c)}+\mathcal{O}(r^3)\,,\label{B3}\\
\tilde{\omega}&=\tilde{\omega}_c-\frac{2}{5}\frac{A_c(\epsilon_c+p_c)r^2}{q^2\eta-4A_c\kappa}\tilde{\omega}_c+\mathcal{O}(r^3)\,,\label{B4}\\
\left(\phi'\right)^{2}&=\frac{p_c}{\eta}r^2-\frac{2q^2(3p_c+\epsilon_c)}{3(3q^2\eta-4\kappa A_c)}r^2+\mathcal{O}(r^3)\,.\label{B5}
\end{align}
\end{subequations}
where the subscript ``$c$'' means that the various variables are
evaluated at the center of the star. Following \cite{Cisterna:2015yla}
we set $A_{c}
=1$
and chose $\tilde{\omega}_{c} = 1$.  Given a choice of EOS, the
central pressure $p_c$ uniquely determines a NS model.

From these expansions we can obtain constraints that must be satisfied by $q^2 \eta$ to
obtain physically acceptable solutions. If we demand that
$p''(r) < 0$~\cite{Delgaty:1998uy}, we obtain
\begin{equation}
q^2 \eta < \frac{4\kappa}{3}\,,
\label{eq:cond1}
\end{equation}
which is automatically satisfied when $\eta < 0$, but sets an upper
bound on $q^2 \eta$ when $\eta > 0$.
On the other hand, the requirement that the derivative of the scalar
field should be real, i.e., $(\phi')^2 > 0$, implies
\begin{equation}
\frac{p_c}{\eta} - \frac{2q^2(3p_c+\epsilon_c)}{3(3q^2\eta-4\kappa)} > 0\,.
\label{eq:cond2}
\end{equation}
For $\eta > 0$ this condition is always satisfied by virtue of
Eq.~(\ref{eq:cond1}). However, when $\eta < 0$ we obtain a lower
bound on $q^2 \left\vert \eta \right\vert$, namely,
\begin{equation}
q^2 \left\vert \eta \right\vert > \frac{3}{4 \pi}\frac{p_c}{2\epsilon_c - 3 p_c}\,.
\end{equation}

To construct NS models we integrate the system of equations
\eqref{ODE1}-\eqref{ODE3}, supplemented by the boundary conditions
\eqref{B1}-\eqref{B3}, from $r=0$ up to the star's radius $r=R$, which
corresponds to the point where the pressure vanishes,
i.e., $p(R)=0$. Then we match the interior solution to the exterior
Schwarzschild metric. The NS mass is obtained by solving the system
\begin{equation}
A(R)=A_{\infty}\left(1-\frac{2M}{R}\right) \,,\quad
A'(R)=A_{\infty}\frac{2M}{R^2}\,,
\label{eq:ainf}
\end{equation}
where $A_\infty$ is an integration constant. Then we rescale the time
variable ($t\rightarrow t\sqrt{A_\infty}$) so that it represents the
coordinate time measured by an observer at infinity.  Because of the
linear dependence of the scalar field on $t$, we correspondingly
rescale $q$ as
\begin{equation}
\label{qinfAinf}
q_\infty=\frac{q}{\sqrt{A_\infty}}\,.
\end{equation}
The stellar structure equations depend only on the combination
$q^2\eta$, so we can set $\eta=\pm 1$ without loss of generality. The
scalar field is computed from Eq.~\eqref{ODE5} for families of
solutions with fixed values of $ q_{\infty}$.  To facilitate
comparisons with~\cite{Cisterna:2015yla}, here we choose these values
to be 0, 0.032, and 0.064. To obtain the solutions we apply a shooting
method, adjusting the value $q$ in each integration until we obtain
the desired value of $q_{\infty}$.

We also integrate Eq.~\eqref{ODE4} for a given $\tilde{\omega}_c$ and we
compute the star's angular velocity $\Omega$ and its angular momentum
$J$, requiring that at the surface
\begin{equation}
\tilde{\omega}(R)=\Omega-\frac{2J}{R^3} \,, \quad
\tilde{\omega}'(R)=\frac{6J}{R^4}\,.
\end{equation}
The moment of inertia is computed through $I = J/\Omega$.
We note that rescaling $\tilde{\omega}(r)$ by a constant factor does
not affect Eq.~\eqref{ODE4}. Therefore, once the solution
$\tilde{\omega}_\tn{old}$ has been found for given initial conditions,
yielding a value $\Omega_\tn{old}$, a new solution
$\tilde{\omega}_\tn{new}$ can immediately be found via
$\tilde{\omega}_\tn{new}=
\tilde{\omega}_\tn{old}\Omega_\tn{new}/\Omega_\tn{old}$.
The moment of inertia $I$ is independent of the star's angular velocity.

In Fig.~\ref{fig:massradius} we show the mass-radius diagram for all
the EOS models used in this paper.
The polytropic case (bottom-right panel) matches the results
in~\cite{Cisterna:2015yla}, except for what we believe to be a
mislabeling of some curves in their Fig.~2.

As pointed out in \cite{Cisterna:2015yla}, the limit
$q_\infty \rightarrow 0$ does not correspond to GR, and indeed the
corresponding mass-radius curves are different from those of GR (solid
black lines). For any EOS and fixed $q_\infty$, positive (negative)
values of $\eta$ correspond to more (less) compact configurations. At
fixed $\eta$, larger values of the scalar charge $q_\infty$
corresponds to stellar models with larger radii. As a reference, the
horizontal colored band correspond to the most massive known NS, PSR
J0348+0432, with $M = 2.01\pm0.04\,
M_{\odot}$~\cite{Demorest:2010bx}.
When $\eta>0$, for all values of $q_{\infty}$ and EOS models
considered in this paper such massive NSs are not supported.

In Fig.~\ref{fig:inertiamass} we show the moment of inertia as a
function of mass for the same stellar models and theory parameters as
in Fig.~\ref{fig:massradius}. In addition, in Table~\ref{tableI14} we
list the values of $I$ for a canonical NS with mass
$M=1.4\,M_\odot$. It is interesting that some theories with $\eta>0$
cannot support stars with this value of the mass.  As expected,
deviations with respect to GR grow as the scalar charge increases,
yielding larger (smaller) moments of inertia for $\eta<0$ ($\eta>0$).
The relative deviation from GR can be of order $~30\%$ for
$q_\infty=0.064$ and $\eta=-1$.

\begin{table}[t]
\centering
\begin{tabular}{ccccc}
\hline
\hline
$\eta$ &
$q_\infty$
& $I_\tn{APR}$
& $I_\tn{GNH3}$
& $I_\tn{SLy4}$ \\
& & ($10^{45}$g cm$^2$) & ($10^{45}$g cm$^2$) & ($10^{45}$g cm$^2$)  \\
\hline
- &GR
&1.31 
&1.81 
&1.37 
\\
- & 0
& 1.28 
& 1.80 
& 1.35 
\\
-1 & 0.032
&1.39 
&1.96 
&1.47 
\\
-1 & 0.064
& 1.70 
& 2.42 
& 1.81 
\\
1 & 0.032
& 1.17  
& 1.64 
& 1.22 
\\
1 & 0.064 & -  & - & - \\
\hline
\hline
\end{tabular}
\caption{Moment of inertia for a NS with $M=1.4\,M_\odot$ for selected
  values of $q_\infty$ and for nuclear-physics-motivated EOS models.
  For $q_{\infty} = 0.064$ and $\eta = 1$, none of the EOS models
  considered here supports NSs with $M=1.4\,M_{\odot}$.}
\label{tableI14}
\end{table}

In GR, the dimensionless moment of inertia $\bar{I} \equiv I/M^3$ was
recently shown to be related to the NS compactness ${\cal C}$ by a
universal relation which is almost insensitive to the adopted
EOS~\cite{Breu:2016ufb}
(see~\cite{1994ApJ...424..846R,Lattimer:2000nx,Bejger:2002ty,Lattimer:2004nj}
for earlier studies):
\begin{equation}
\bar{I}_\tn{fit}=a_1{\cal C}^{-1}+a_2{\cal C}^{-2}+a_3{\cal C}^{-3}
+a_4{\cal C}^{-4}\ ,\label{fit1}
\end{equation}
where the fitting coefficients $a_{i}$, $i = (1,\dots,4)$, are listed
in Table II of \cite{Breu:2016ufb}. This $I$-${\cal C}$ relation
reproduces numerical results with an accuracy better than $3\%$. The
observed universality is reminiscent of the $I$-Love-$Q$ relations
between the moment of inertia, tidal deformability (as encoded in the
so-called Love number) and rotational quadrupole moment
$Q$~\cite{Yagi:2013bca,Yagi:2016ejg}. The extension of these
near-universal relations $I$-${\cal C}$ relations to theories of the
Ringo and John subclasses is discussed in the Appendix.

It is natural to ask whether these stealth NS models are stable.
Vacuum, static, spherically symmetric solutions where the scalar field
has a linear time dependence were shown to be free from ghost and
gradient instabilities under odd-parity gravitational perturbations as
long as the following conditions are met~\cite{Ogawa:2015pea}:
\begin{equation}
{\cal F}>0, \quad {\cal G}>0,\quad {\cal H}>0,
\end{equation}
where
\begin{align}
\label{calf}
{\cal F}&=
2\left(G_4-\frac{q^2}{A} G_{4X}\right)
=
2\left(\kappa+\frac{q_\infty^2\eta}{4}-\frac{q_\infty^2\eta A_\infty}{2A}\right)\,,
\nonumber \\
\\
\label{calg}
{\cal G}&=2\left(G_4-2XG_{4X}+\frac{q^2}{A} G_{4X}\right)
\nonumber\\
&=2\left(\kappa-\frac{q_\infty^2\eta}{4}+\frac{q_\infty^2\eta A_\infty}{2A}\right)\,,
\\
\label{calh}
{\cal H}&=2 \left(G_4-2XG_{4X}\right)
=
2\left(\kappa-\frac{q_\infty^2\eta}{4}\right)\,.
\end{align}
Here we have used $X=q^2/(2A_\infty)$ as well as Eq.~\eqref{qinfAinf},
which applies to the stealth BH solutions of~\cite{Babichev:2013cya}.
For stealth BH solutions, $A\to 0$ in the vicinity of the event
horizon; therefore the third term on the right-hand side of
Eqs.~\eqref{calf} and \eqref{calg} is the dominant one.  As a
consequence ${\cal F}{\cal G}<0$, suggesting that these solutions are
generically unstable~\cite{Ogawa:2015pea}.

A similar argument can be applied to our stealth NS solutions.  In the
exterior vacuum spacetime of the star, the metric function $A$, which
satisfies $A < A_\infty$, remains positive and finite.  When $\eta$ is
{\it positive}, ${\cal G}$ is always positive as well, and the
conditions ${\cal F}>0$ and ${\cal H}>0$ everywhere outside the star
translate into
\begin{eqnarray}
q_\infty^2\eta&<&4\kappa\frac{A(R)}{2A_\infty-A(R)} =
4\kappa\, \left(\frac{1-2{\cal C}}{1+2{\cal C}}\right)\,,
\label{eq:nsF}
\\
q_\infty^2\eta&<&4\kappa \,,
\label{eq:nsH}
\end{eqnarray}
respectively, where we have used Eq.~(\ref{eq:ainf}).

We have numerically confirmed that all NS models presented in
Fig.~\ref{fig:massradius} satisfy the conditions \eqref{eq:nsF} and
\eqref{eq:nsH} for the largest value of $q_\infty=0.064$ considered in
this paper.
For a typical NS the compactness is ${\cal C} \approx 0.2$, and the
right-hand side of Eq.~\eqref{eq:nsF} is approximately $0.035$, which
is much larger than our choice $q_0^2\eta = 0.064^2\approx 0.004$.
The condition (\ref{eq:nsF}) will be violated only for an
unrealistically compact NS with ${\cal C} \approx 0.45$.  This
suggests that hypothetical ultracompact objects -- such as Lemaitre
stars~\cite{BowersLiang:1974,Glampedakis:2013jya,Yagi:2016ejg} and
gravastars~\cite{Pani:2009ss,Cardoso:2014sna,Pani:2015tga} -- may be
unstable in the presence of a stealth scalar field.

Similarly, for {\it negative} values of $\eta$, ${\cal F}$ and
${\cal H}$ are always positive,
and the condition ${\cal G}>0$ is satisfied everywhere outside the star
if
\begin{eqnarray}
\label{eq:nsG}
q_\infty^2|\eta|<4\kappa\frac{A(R)}{2A_\infty-A(R)} = 4\kappa\, \left(\frac{1-2{\cal C}}{1+2{\cal C}}\right)\,.
\end{eqnarray}
We have also checked that for $q_\infty=0.064$ and $\eta=-1$,
all NS models presented in Fig.~\ref{fig:massradius} satisfy \eqref{eq:nsG}.
In the Newtonian limit ${\cal C} \ll 1$, the stealth NS spacetime is
stable for $q_\infty^2\eta <4\kappa$ when $\eta>0$, and for
$q_\infty^2|\eta|<4\kappa$ when $\eta<0$. For NSs with larger values
of $q_\infty^2|\eta|$ the exterior spacetime becomes unstable
everywhere, including the Newtonian regime.

It is interesting to consider the nonrelativistic limit of theories
of the John class.  Introducing the usual mass function $m(r)$ such
that $B(r)=1-2m(r)/r$, we see that the pressure equation retains its
standard form
\begin{equation}
\frac{dp}{dr} = -\frac{m \rho}{r^2}\,,
\label{eq:dpdrNewton}
\end{equation}
where $\rho$ is the mass density.
However the mass equation is reduced to
%
\begin{equation}
\frac{dm}{dr} = \frac{4\pi r^2 \rho}{1 - 12\pi q^2 \eta}\,.
\label{eq:dmdrNewton}
\end{equation}
This behavior looks reminiscent of ``beyond Horndeski''
theories~\cite{Gleyzes:2014dya,Gleyzes:2014qga}, where a partial
breakdown of the Vainshtein mechanism occurs, modifying the Newtonian
limit~\cite{Kobayashi:2014ida}.  In fact, several authors have
advocated the use of this ``feature'' to constrain beyond Horndeski
theories using Newtonian stars or white
dwarfs~\cite{Sakstein:2015aac,Sakstein:2015zoa,Koyama:2015oma,Jain:2015edg,Saito:2015fza}.
While those theories modify the pressure
equation~(\ref{eq:dpdrNewton}), leaving the mass equation unaltered,
theories of the John subclass seem to alter the Newtonian limit in the
opposite way.

However, combining Eqs.~(\ref{eq:dpdrNewton})-(\ref{eq:dmdrNewton})
and restoring the gravitational constant $G$ we obtain
\begin{equation}
\frac{1}{r^2} \frac{d}{dr} \left( \frac{r^2}{\rho} \frac{dp}{dr}\right) =
-4 \pi {G}_{\rm eff} \rho\,,
\end{equation}
which is equivalent to the ordinary hydrostatic equilibrium equation
in Newtonian gravity with an {\it effective} gravitational constant
\begin{equation}
{G}_{\rm eff} \equiv \frac{G}{1-12\pi q^2\eta}\,.
\label{eq:effectiveG}
\end{equation}
Therefore the nonrelativistic limit of the ``John'' theories
considered in this section is equivalent to Newtonian gravity with an
effective gravitational constant $G_{\rm eff}$. Incidentally, a
similar result was found by Cisterna {\it et al.}~\cite{Cisterna:2016vdx} in
the context of cosmology [cf. their Eq.~(38)].


\subsection{Paul\\(Double-dual of the Riemann tensor)}
\label{sec:paul}
Let us now turn to NS solutions in theories containing the Paul
Lagrangian~\eqref{paul}. We start with the simplest model, given by
the combination
\begin{align}
{\cal L}&={\cal L}_{\rm george}+{\cal L}_{\rm paul}\nonumber\\
&=
R-\frac{1}{3}\alpha P^{\mu\nu\alpha\beta}\nabla_\mu\phi\nabla_\alpha\phi\nabla_\nu\nabla_\beta\phi
\,,\label{paulaction}
\end{align}
which from Eqs.~\eqref{horndeski_fab4} corresponds to the following
choice of the functions $G_{i}$:
\begin{equation}
G_2=G_3=0\,,\quad G_4=1\,,\quad
G_5=
\alpha
 X\,,
\end{equation}
where $\alpha$ is a coupling parameter.  As in Sec.~\ref{sec:nonmin},
we consider a scalar field with linear time dependence of the
form~\eqref{ansatzphi}. This choice is crucial for $\phi(r)$ to have a
nontrivial profile. Indeed, the nonvanishing components of the scalar
current for the action~\eqref{paulaction} are
\begin{align}
J^r &= \frac{\alpha}{2r^2}\frac{B}{A}\left[q^2(B-1)+A(1-3B)B\phi'^2\right]\frac{A'}{A}\,,\label{JrPaul}\\
J^t &= \frac{q\alpha}{2r^2}\frac{B}{A}\bigg\{\phi'\left[\frac{A'}{A}(B-1)+\frac{B'}{B}(3B-1)\right]\nonumber\\
&+2(B-1)\phi''\bigg\}\,.
\end{align}
From the first equation we conclude that in the limit $q\rightarrow0$
the condition $J^r=0$ implies $\phi'=0$; i.e., the scalar field must
be constant. However for $q\neq0$ we obtain
\begin{equation}
(\phi')^2 = q^2 {\frac{1-B}{A(1-3B)B}}\,.
\label{eq1paul}
\end{equation}
Replacing this relation into the $(tt)$ and $(rr)$ components of
Eqs.~\eqref{EoM}, we derive two first-order equations for the metric
variables $A$ and $B$:
\begin{align}
B'&=\frac{1-B-8\pi r^2 \epsilon }{r-\frac{q^3\alpha\sqrt{1-B}[AB(1-3B)]^{3/2}}{A^3(1-3B)^3}}\ ,\label{eq2paul}\\
A'&=\frac{A^3}{B}\frac{1-B+8\pi r^2p}{A^2r-\frac{q^3\alpha B\sqrt{1-B}\sqrt{AB(1-3B)}}{(1-3B)^2}}
\label{eq3paul}\ .
\end{align}
Equations~\eqref{eq1paul}-\eqref{eq3paul}, together with a choice of EOS and
the energy-momentum conservation equation~\eqref{enecons}, which gives
\begin{equation}
p'=-\frac{\epsilon+p}{2}\frac{A'}{A}\,,\\
\end{equation}
form a closed system of differential equations, which can be
integrated by imposing suitable initial conditions at the center of
the star. These conditions can be found through a Taylor expansion in
$r$,
\begin{subequations}
\begin{align}
A(r)&=A_c+\frac{r^2}{q^6\alpha^2}A_2(p_c,A_c) + {\cal O}(r^3)\,,\label{eq2pauIni}\\
B(r)&=1+\frac{r^2}{q^6\alpha^2}B_2(p_c,A_c) + {\cal O}(r^3)\,,\label{eq3pauIni}\\
p(r)&=p_c+\frac{r^2}{q^6\alpha^2}p_2(p_c,A_c) + {\cal O}(r^3)\,,\label{eq5pauIni}\\
\phi'(r) &= \pm \sqrt{\frac{B_2(p_c,A_c)}{2A_c}} \frac{r}{q^2 \alpha} + {\cal O}(r^3),
\end{align}
\end{subequations}
where $A_2$, $B_2$, and $p_2$ are functions of the constant parameters
$A_c$ and $p_c$.  Unlike Eqs.~\eqref{eq2paul} and \eqref{eq3paul},
which reduce to GR for $\alpha\rightarrow0$ (or $q\rightarrow 0$), the
initial conditions for the metric functions, $(\phi')^2$, and the
pressure are ill defined.
Note that such a pathological behavior is not expected in the naive $\alpha\to 0$ limit of \eqref{eq2paul} and \eqref{eq3paul}, because this is a ``nonperturbative'' effect such that the leading behavior $\sqrt{1-B}\propto 1/\alpha$ obtained from \eqref{eq3pauIni} cancels the $\alpha$ terms in \eqref{eq2paul} and \eqref{eq3paul},
making the deviation from GR evident.

To better understand this issue, let us reconsider the $\eta\to 0$
limit of the John action.  In that case, as we see from
Eqs.~\eqref{B1}-\eqref{B5}, the only divergent quantity as $\eta\to 0$
is the derivative of the scalar field $\phi'$, while all other metric
and matter quantities have a finite limit.  Since we work in the
Jordan frame there is no direct coupling between the scalar field and
matter. Furthermore the scalar field does not backreact on the
spacetime in the stealth exterior, and therefore a singular
behavior of the scalar field does not affect the geodesics of
particles outside the star.  In contrast, for the Paul case all
physical quantities diverge in the limit $\alpha\to 0$, indicating a
pathological behavior in the NS interior. Furthermore, at variance
with the John case, we could not find a stealth exterior solution for
Paul.  Our results suggest that exterior stealth solutions for Paul do
not exist under the ansatz~\eqref{ansatzphi} for the scalar field.

We observed a similar behavior for other Fab Four theories
involving the Paul term. We considered the following
combinations:
\begin{align}
&{\cal L}_{\rm george}+{\cal L}_{\rm paul}+{\cal L}_\tn{K}\ ,\quad\phi(r)\ ,\\
&{\cal L}_{\rm george}+{\cal L}_{\rm paul}+{\cal L}_{\rm john}+{\cal L}_\tn{K}\ ,\quad\phi(t,r)\ ,\\
&{\cal L}_{\rm george}+{\cal L}_{\rm paul}+{\cal L}_{\rm ringo}+{\cal L}_\tn{K}\ ,\quad\phi(t,r)\ ,\\
&{\cal L}_{\rm george}+{\cal L}_{\rm paul}+{\cal L}_{\rm john}+{\cal L}_{\rm ringo}+{\cal L}_\tn{K}\ ,\quad \phi(t,r)\ .
\end{align}
In all of these cases the physical variables suffer from the same divergence
when the coupling parameter $\alpha$ of the Paul term vanishes.

Appleby~\cite{Appleby:2015ysa} found that the self-tuning mechanism is
not applicable for spherically symmetric black hole spacetimes in
theories of the Paul class. Our results strengthen his
conclusions, suggesting that the Paul term does not allow for
physically well-behaved compact object solutions.


\section{Conclusions}
\label{sec:conclusions}

We have presented an exhaustive study of slowly rotating NS solutions
in the shift-symmetric class of Fab Four gravity, namely, the
subclass of Horndeski's gravity that may allow for dynamical
self-tuning of the quantum vacuum energy, and for this reason has been
the subject of intense scrutiny in a cosmological context. Our main
goal was to investigate whether Fab Four gravity is compatible with
the existence of relativistic stars, such as NSs.

Among the nonminimal couplings in Fab Four gravity listed in
Eqs.~\eqref{john}-\eqref{ringo}, we especially focused on the John
(nonminimal derivative coupling to the Einstein tensor) and Paul
(nonminimal derivative coupling to the double dual of the Riemann
tensor) subclasses. This is both because George (GR) and Ringo
(EdGB gravity) have been extensively studied in the past and because
Joh and Paul are the crucial terms allowing for self-tuning of
the quantum vacuum energy in cosmological scenarios.

In the case of John, if we make the
assumption that the scalar field has a linear time dependence of the
form~\eqref{ansatzphi}, there is a stealth solution such that the
scalar field does not backreact on the metric in the exterior, while
it introduces nontrivial modifications of the interior stellar
structure with respect to GR in the stellar interior.  Our results on
spherically symmetric NSs agree with previous
work~\cite{Cisterna:2015yla} and extend it to slowly rotating
solutions.  As pointed out in~\cite{Cisterna:2015yla}, in the limit of
vanishing scalar charge ($q_\infty\to 0$) the mass-radius curves
differ from GR.  Irrespective of the chosen EOS, positive (negative) values
of the coupling constant $\eta$ in \eqref{nonminact} yield more (less)
compact stellar configurations.
For positive
values of $\eta$ this fact can be used to put mild (EOS-dependent) constraints
on the maximum value of $q_{\infty}$; cf. Ref.~\cite{Cisterna:2016vdx}.

We have also shown that the approximately EOS-independent relations
between the moment of inertia $I$ and compactness ${\cal C}$ within GR
break down in this theory.  Therefore, in principle, future
measurements of $I$ could potentially constrain the value of
$q_{\infty}$~\cite{Kramer:2009zza}.  We also obtained improved
$I$-${\cal C}$ relations that depend of the value of $q_{\infty}$ and
are accurate within $\sim 5 \%$.

Based on stability studies in the context of BH
solutions~\cite{Ogawa:2015pea}, we have argued that the NS models
studied here are generically stable under odd-parity gravitational
perturbations. A systematic study of stellar perturbations within
theories of the John subclass is desirable, and it could follow in the
footsteps of similar studies for scalar-tensor
theory~\cite{Sotani:2004rq,Sotani:2005qx,Silva:2014ora,Sotani:2014tua} and EdGB
gravity~\cite{Blazquez-Salcedo:2015ets}.

Surprisingly, we also found that in all subclasses of the Fab Four and its
minimal extensions that involve Paul, not only the scalar field, but
also all metric functions and the pressure diverge at the center of
the star in the small-coupling limit. Therefore ``healthy'' BH and
stellar solutions do not seem to exist in the shift-symmetric Paul
subclass. It will be interesting to determine whether this conclusion
still holds in the absence of shift symmetry.

As a straightforward generalization of the present work, one could
search for NS solutions in Fab Four theories where the potentials
\eqref{john}-\eqref{ringo} have nontrivial functional forms, as well
as in more general (non-shift-symmetric) versions of Horndeski's
theory.  The general formalism developed in~\cite{Maselli:2015yva} can
be straightforwardly applied to these cases.

Barausse and Yagi~\cite{Barausse:2015wia} have recently shown that the
so-called sensitivities of compact objects~\cite{Eardley:1975} vanish
in shift-symmetric Horndeski gravity, which includes the Fab Four
class. Consequently the dynamics of binaries involving NSs is, to
leading post-Newtonian order, the same as in GR. It would be
interesting to determine whether these conclusions hold at higher
post-Newtonian orders, and whether gravitational waves can be used at
all to constrain these theories.

Recently, a similar study of slowly rotating
stars appeared on the arXiv~\cite{Cisterna:2016vdx}. Their work
focuses on theories of the John class and deals also with their
cosmological interpretation. Where our works overlap, they agree with
our main conclusions.

\acknowledgments

We thank Massimiliano Rinaldi, T\'erence Delsate, Jeremy Sakstein,
Kazuya Koyama Eugeny Babichev, and Thomas Sotiriou
for discussions and correspondence. H.O.S was supported by NSF CAREER Grant
No. PHY-1055103.  E.B. was supported by NSF CAREER Grant
No. PHY-1055103 and by FCT Contract No. IF/00797/2014/CP1214/CT0012 under
the IF2014 Programme.  M.M. was supported by the FCT-Portugal through
Grant No. SFRH/BPD/88299/2012. This work was also supported by 
the H2020-MSCA-RISE-2015 Grant No. StronGrHEP-690904

\appendix

\section{$I$-${\cal C}$ relations}
\label{app:IC}

This appendix discusses the relation between the moment of inertia
and the compactness for NSs in theories of the John and Ringo
subclasses.

\subsection{John\\(Nonminimal coupling with the Einstein tensor)}

\begin{figure}[t]
\includegraphics[width=\columnwidth]{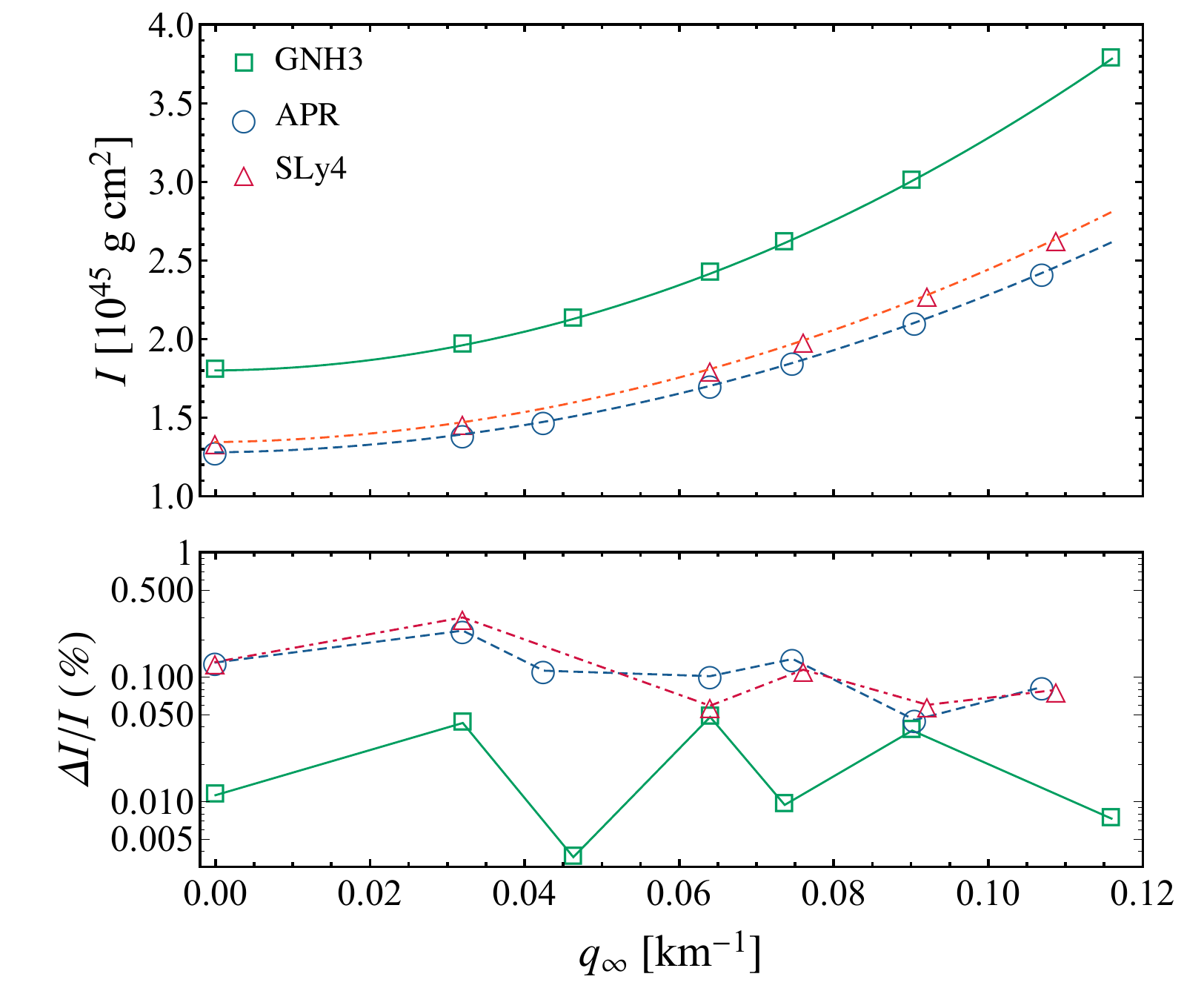} \quad
\caption{Top panel: The moment of inertia $I$ versus charge $q_\infty$
for a canonical NS with $M=1.4\,M_\odot$ and $\eta=-1$,
and the realistic EOSs APR, GNH3, and SLy4.
Bottom panel: Relative percentage errors between the numerical
data and the relation \eqref{fitq}.}
\label{fig:inertiaq}
\end{figure}

The behavior of $I$ as function of $q_\infty$ can be
accurately described by a simple quadratic fit of the form
\begin{equation}
I=p_0+p_1 q_\infty+p_2q_\infty^2\ ,\label{fitq}
\end{equation}
where $(p_0,\,p_1,\,p_2)$ are constants.  In the top panel of
Fig.~\ref{fig:inertiaq} we compare this relation with numerical data
for $\eta=-1$ (note that for this figure we have computed models with
additional values of $q_\infty$ that were not displayed in
Figs.~\ref{fig:massradius} and \ref{fig:inertiamass} to avoid
cluttering). The bottom panel of Fig.~\ref{fig:inertiaq} shows that
the relative errors between the numerical data and the fit are
typically of order $0.1\%$ or smaller.

\begin{figure}[tbh]
\centering
\includegraphics[width=\columnwidth]{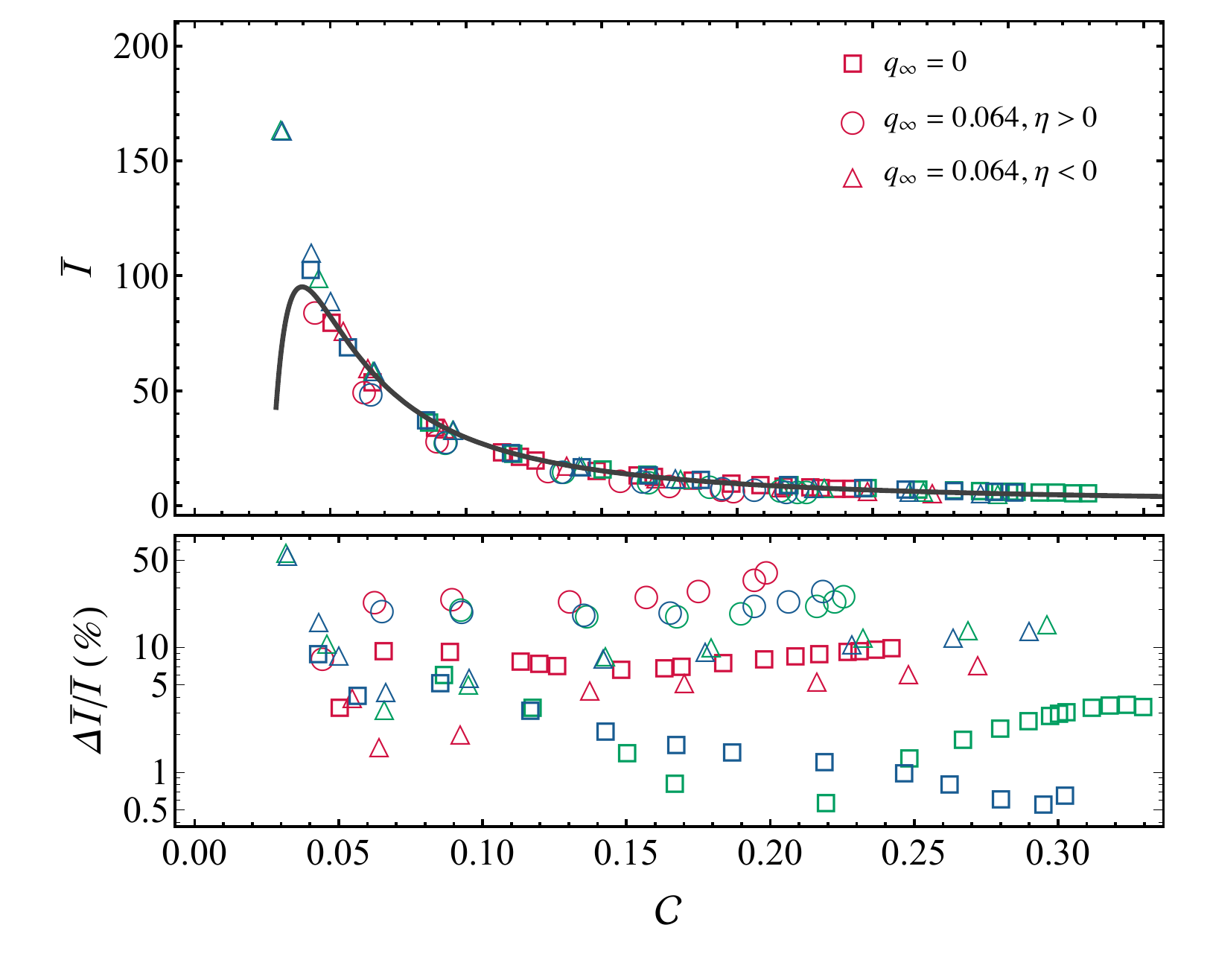}
\caption{Top panel: $\bar{I}$-${\cal C}$ relation for different values of
  the scalar charge $q_\infty$ and the realistic EOS APR (blue), GNH3
  (red), SLy4 (green). The solid curve represents the fit given
  by Eq.~\eqref{fit1}, obtained in~\cite{Breu:2016ufb}.
  Bottom panel: Relative errors between the numerical
  data and the analytic relation. For illustrative purposes, we show
  the cases $q_{\infty} = 0$ and $q_{\infty} = 0.064$. For the
  latter, deviations from GR are more dramatic.}
\label{fig:relativerror}
\end{figure}

To understand whether these relations hold also for theories of the
John subclass, we have compared our numerical data against
Eq.~\eqref{fit1}, computing the relative error
$\Delta \bar{I}/\bar{I}=\vert 1-\bar{I}_\tn{fit}/\bar{I}\vert$. The
results are shown in the bottom panel of
Fig.~\ref{fig:relativerror}. Errors are always larger than in GR, and
they can be as high as 40\% for low-compactness configurations. A
similar trend is observed for the $I$-Love-$Q$ relations in GR
in~\cite{Silva:2016myw}.  Deviations from the GR relation are due to
the strong dependence of the star's bulk properties on the scalar
charge $q_\infty$, which spoils the (approximate) EOS universality of
the relation proposed in~\cite{Breu:2016ufb}. Therefore we conclude
that a theory-independent fit would perform poorly.

\begin{figure}[t]
\centering
\includegraphics[width=\columnwidth]{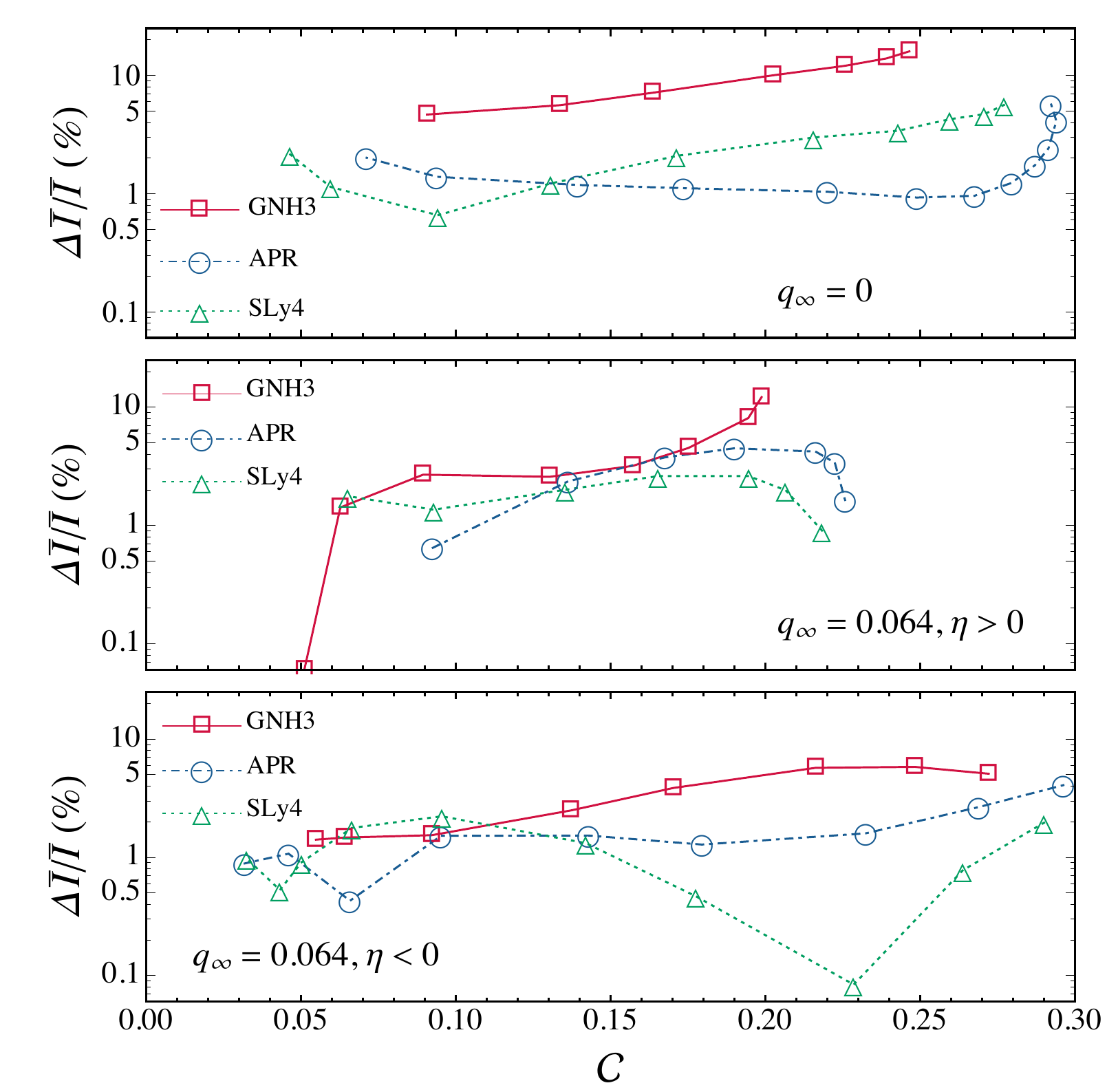}
\caption{Relative errors between the improved fits and the numerical data.
  Top panel: $q_\infty=0$. Middle panel: $q_\infty=0.064$ and
  $\eta=1$. Lower panel: $q_\infty=0.064$ and
  $\eta=-1$.}
\label{fig:relativerrornew}
\end{figure}

It is still possible to introduce approximately EOS-independent
relations for $I$-${\cal C}$ at fixed values of the theory parameters
$q_{\infty}$ and $\eta$ using the functional form given in
Eq.~\eqref{fit1}. The relative errors between the numerical data and
these fits are shown in Fig.~\ref{fig:relativerrornew}, and the
corresponding fitting coefficients are listed in Table~\ref{tablefit}.
For almost all configurations the new relations perform better than
Eq.~\eqref{fit1}, with relative errors that can be an order of
magnitude smaller.

\subsection{Ringo\\(Einstein-dilaton-Gauss-Bonnet gravity)}

We have also investigated the $I$-${\cal C}$ relations for theories of
the Ringo subclass (EdGB gravity) using the numerical data
from~\cite{Pani:2011xm}. We found that the fit proposed
in~\cite{Breu:2016ufb} works remarkably well for EdGB, with relative
percentage errors $\lesssim 10\%$ for a wide range of
compactness. This result is complementary to the $I$-$Q$ relations in
EdGB obtained in~\cite{Kleihaus:2014lba}. We recall, however, that our
calculations are limited to slow rotation. The question of whether or
not rapidly rotating NSs in EdGB satisfy the same $I$-${\cal C}$
relations of~\cite{Breu:2016ufb} could be addressed following the
analysis of~\cite{Kleihaus:2014lba,Kleihaus:2016dui}.

\begin{table}[b]
\centering
\begin{tabular}{cccccc}
\hline
\hline
$q_\infty$ & $\eta$ & $a_1$ & $a_2$ & $a_3$ & $a_4$\\
\hline
0 & $-$ & 0.684 & 0.265 & $-0.0062$ & $6.87 \times 10^{-5}$ \\
0.032 & $1$ & 0.666 & 0.240 & $-0.00364$ & $-2.01\times 10^{-6}$\\
0.032 & $-1$ & 0.776 & 0.273 & $-0.00809$ & $1.64 \times 10^{-4}$ \\
0.064 & $1$ & 0.0654 & 0.348 & $-0.0125$ & $1.81 \times 10^{-4}$ \\
0.064 & $-1$ & 0.872 & 0.276 & $-0.00574$ & $4.53 \times 10^{-5}$ \\
\hline
\hline
\end{tabular}
\caption{Numerical coefficients of the new universal $I$-${\cal C}$
relations, for fixed values of $q_\infty$ and $\eta$.}
\label{tablefit}
\end{table}

\newpage

\bibliography{bibnote}
\end{document}